\documentclass[prd,aps,showpacs,epsf,floats,onecolumn]{revtex4}
\usepackage[utf8x]{inputenc}
\usepackage{graphicx}
\usepackage{amsfonts}
\usepackage{amsmath}
\usepackage{amssymb}
\usepackage{hyperref}
\usepackage{bookmark}
\usepackage{xcolor}
\usepackage{soul}
\allowdisplaybreaks

\newcommand{\be}[1]{\begin{equation}\label{#1}}
\newcommand{\ee}{\end{equation}}
\newcommand{\bea}[1]{\begin{eqnarray}\label{#1}}
\newcommand{\eea}{\end{eqnarray}}
\newcommand{\no}{\nonumber \\}
\newcommand{\Fig}[1]{Fig.(\ref{#1})}

\newcommand{\Eq}[1]{Eq.(\ref{#1})}

\newcommand{\bq}{\begin{quote}}
\newcommand{\eq}{\end{quote}}
\newcommand{\benum}{\begin{enumerate}}
\newcommand{\eenum}{\end{enumerate}}
\newcommand{\bpm}{\begin{pmatrix}}
\newcommand{\epm}{\end{pmatrix}}
\newcommand{\bsub}{\begin{subequations}}
\newcommand{\esub}{\end{subequations}}
\newcommand{\bwt}{\begin{widetext}}
\newcommand{\ewt}{\end{widetext}}

\usepackage{color}

\def\trm#1{\textrm{#1}}
\def\tit#1{\textit{#1}}

\def\defn{\overset{\textrm{def}}{=}}
\def\bk{\bigskip}

\def\a0{{\alpha_0}}

\def\da0{{\dot{\alpha}_0}}

\def\myoverDefn#1#2{\hbox{\space \raise-2mm\hbox{$\textstyle{#1} \atop \scriptstyle{#2}$} }}

\def\mO{{\mathcal{O}}}

\def\d{{\delta}}
\def\a{{\alpha}}

\def\dag{\dagger}

\def\rp{r_{P}}
\def\rp2{r_{p}^{2}}
\def\Tr{\textrm{Tr}}

\newcommand{\half}{\frac{1}{2}}
\newcommand{\thalf}{\tfrac{1}{2}}

\newcommand{\ket}[1]{|#1\rangle}
\newcommand{\bra}[1]{\langle #1|}
\newcommand{\pd}{\partial}
\newcommand{\avg}[1]{\langle {#1}\rangle}

\newcommand{\EV}[3]{\langle {#1} | {#2}| {#3} \rangle}

\def\cov{\trm{cov}}

\def\DA{{\Delta A}}
\def\DB{{\Delta B}}
\def\ev#1{{\langle #1 \rangle}}

\begin{document}

\title{Information-Geometric Representation of Upper Bounds on \\Fluctuation Growth in Open Quantum Systems}

%

\author{\textbf{Newshaw Bahreyni}}
\email{corresponding author: newshaw.bahreyni@pomona.edu}
\affiliation{Pomona College, Claremont, CA 91711, USA}
\author{\textbf{Paul M. Alsing}}
\email{alsingpm@gmail.com; palsing@albany.edu}
\affiliation{University at Albany-SUNY, Albany, NY 12222, USA}

\begin{abstract}
Building on our earlier results for closed quantum systems, we investigate the dynamical evolution of quantum fluctuations and their connections to Fisher information and the quantum operations describing open-system dynamics.  Starting from a general expression for the time derivative of the uncertainty of an observable, we derive an alternative expression for the upper bound on fluctuation growth in an open quantum system, which was found in our previous work, using the Cauchy–Schwarz inequality.  The resulting bound separates contributions arising from the evolution of the quantum state and from explicit time dependence of the observable, and reduces to a known result for closed-system dynamics.  We then establish a connection between fluctuation dynamics and quantum Fisher information by decomposing the density operator and the symmetric logarithmic derivative (SLD) into coherent and incoherent contributions. This leads to a bound relating the evolution of observable uncertainty to the coherent and incoherent components of the Fisher information, providing an information-geometric interpretation of fluctuation growth.

To find the relation between fluctuation dynamics and quantum operators, we derive the time evolution of the eigenvalues of the density operator in terms of Kraus operators. Then by considering the infinitesimal-time relation between Kraus and Lindblad operators, we obtain a corresponding expression for the eigenvalue dynamics in terms of Lindblad operators.  The resulting formulation directly connects changes in the eigenvalues of the density operator to the quantum  processes acting on the system.  We then study the results using some examples.   Our results provide a framework connecting observable fluctuation growth, Fisher information, and the Kraus and Lindblad descriptions of quantum evolution, offering a route toward understanding how specific quantum channels influence the information and fluctuations available in an evolving quantum system.
   
\end{abstract}

\pacs{Quantum Computation (03.67.Lx), Quantum
Information (03.67.Ac), Quantum Mechanics (03.65.-w)}

\maketitle
\section{Introduction}
Fundamental limits on the evolution of quantum systems, known an Quantum Speed Limits (QSLs), have attracted considerable attention in quantum information science, quantum thermodynamics, and quantum control \cite{CMCF, ZHX, EW, MSZ, DC}. Quantum Speed Limits (QSLs) provide lower bounds on the minimum time required for a quantum system to evolve between distinguishable states.  Originating from the Mandelstam–Tamm \cite{MT} and Margolus–Levitin bounds \cite{ML}, QSLs establish a deep connection between the geometry of quantum state space and the generators of quantum evolution, such as the Hamiltonians \cite{DC}.  The geometry tells us how distinguishable two states are, while the generator of evolution tells us how fast the state moves through that geometry. 
 
More recently, these concepts have been extended beyond the characterization of state evolution to encompass the dynamics of physical observables, thereby providing a broader framework for understanding nonequilibrium quantum processes \cite{CRABC}.

While the majority of studies on dynamical bounds have focused on the evolution of expectation values, fluctuation growths of observables form an equally important aspect of quantum dynamics.  The standard deviation of an observable quantifies the spread of measurement outcomes and its time evolution, known as fluctuation growth, provides crucial information in many experimental settings \cite{C, W, SK}.  Fluctuation growths have been explored through establishing universal bounds on fluctuation dynamics, demonstrating that the speed of fluctuation growth is fundamentally limited by the fluctuation of a suitably defined velocity observable.  These results reveal a trade-off between the evolution of the mean value and the fluctuation of an observable, thereby extending the notion of speed limits from expectation values to higher statistical moments \cite{H1, CRABC, BAC}. 

Quantum Acceleration Limits (QALs) characterize the rate of change of the speed of quantum evolution itself.  For unitary dynamics generated by time-dependent Hamiltonians, QALs establish that the temporal variation of the Hamiltonian uncertainty is bounded by the uncertainty of the Hamiltonian’s time derivative.    Building upon this framework, the concept was generalized to arbitrary observables undergoing unitary evolution.  It is shown that the time derivative of the standard deviation of an observable $A$ satisfies an upper bound determined by the standard deviation of an associated velocity-like observable $v_{A}$.  This result provides a direct connection between quantum acceleration limits and fluctuation dynamics, yielding a geometric and operational interpretation of how rapidly observable uncertainties may evolve in closed quantum systems.  Furthermore, the derived inequalities reveal a fundamental trade-off relation between the rates of change of the mean value and the standard deviation of an observable, highlighting the interplay between deterministic evolution and statistical dispersion \cite{P, CCBA}.

The uncertainty of an observable is not only a fundamental measure of quantum fluctuations, but can also be closely related to the sensitivity of a quantum state to changes in an external parameter. This connection is particularly important in quantum metrology, where the quantum Fisher information (QFI) determines the ultimate precision with which a parameter can be estimated \cite{BC}. Understanding how fluctuations evolve in time can therefore provide insight into how quantum information is generated, redistributed, or lost during the evolution of a quantum system.

In this work, we begin with finding an expression for the time evolution of the uncertainty of an observable A,
$2\sigma_A\dot{\sigma}_A
=
\sum_j\dot p_j\langle(\Delta A)^2\rangle_j
+
\left\langle\{\Delta A,\dot A\}\right\rangle$ ,
which separates the evolution of the uncertainty into contributions associated with changes in the state and changes in the observable itself. This relation provides a useful starting point for studying fluctuation growth in both closed and open quantum systems. By applying the Cauchy–Schwarz inequality, we obtain upper bounds on the rate of fluctuation growth and recover the corresponding result for closed-system dynamics as a special case.  This also gives an alternative representation of the  fluctuation-growth inequality for open quantum systems derived in \cite{BAC} that was found to be
\begin{equation}
\label{eq:I1}
\dot{\sigma}^{2}_{A} \leq 2 \left( \left\langle (\partial_{t}A)^{2}\right\rangle +\frac{(\Tr(\dot{\rho}\Delta A^{2}))^{2}}{4\sigma^{2}_{A}}\right),
\end{equation}
where $A$ is an observable and $\rho$ is the density operator.  The first term on the right-hand side describes the contribution arising from the explicit time dependence of the observable, while the second term captures the contribution associated with the evolution of the quantum state and the fluctuations correlated with its dynamical generator.  This reformulation yields an alternative representation of the fluctuation-growth inequality that naturally separates the contributions arising from operator dynamics and probability dynamics.

We then investigate the connection between this fluctuation dynamics and Fisher information. Classical Fisher information characterizes the distinguishability of nearby probability distributions, while quantum Fisher information extends this concept to quantum states and accounts for both changes in the eigenvalues and changes in the eigenvectors of the density operator \cite{FI1, FI2}.  By decomposing the density operator and the symmetric logarithmic derivative into their incoherent and coherent components, we obtain a bound relating the evolution of observable fluctuations to the coherent and incoherent contributions to quantum Fisher information. This establishes a connection between the dynamical growth of fluctuations and the information geometry of the quantum state.

We then focus on the quantum operations responsible for the evolution of the density operator.  In particular, we derive an expression for the time derivative of the eigenvalues $p_{j}$ of the density operator in terms of Kraus operators.  Since Kraus operators provide a general description of completely positive trace preserving (CPTP) quantum channels \cite{NCh}, this formulation establishes a direct connection between the change in the statistical weights of the quantum state and the underlying quantum operation.  We subsequently use the infinitesimal-time relation between Kraus and Lindblad operators to obtain an analogous expression for $\dot{p}_{j}$ in terms of Lindblad operators.

Several examples including amplitude damping, bit flip, phase flip, bit-phase flip, and collective superradiant decay are considered to demonstrate how the derived expression describes probability transfer between quantum states. These examples illustrate that different environmental processes affect the eigenvalues and coherences of the density operator in fundamentally different ways. In particular, processes such as phase flips can preserve the populations while degrading phase information, whereas dissipative processes can produce irreversible population transfer.

Taken together, these results provide a framework connecting observable fluctuations, Fisher information, and the quantum dynamical maps governing open-system evolution. This framework allows the fluctuation-growth bound to be expressed in terms of quantities determined by the state dynamics and, ultimately, by the Kraus and Lindblad operators describing the physical process.

The rest of the paper is structured as follows.  In Section II we will derive an expression for the time evolution of the uncertainty of an observable $A$, then find an alternative representation of Eq. (\ref{eq:I1}).  We will then find the relation between the time evolution of the uncertainty of $A$ and the Fisher information in Section III.  To find the relation between the time evolution of the uncertainty of $A$ and quantum operators, we will find the evolution of the eigenvalues of the density operator in terms of Kraus operators in Section IV, and in terms of Lindblad operators in Section V, followed by some examples in Section VI.  Our concluding remarks will be in Section VII.

\section{Open System Fluctuation Limit}
Upper bounds on fluctuations in an open quantum system have been investigated in \cite{BAC}.  Fluctuation bounds in an open system with and without information about the evolution operator yield the same result given by Eq. (\ref{eq:I1}).  Here we are seeking an alternative representation of the same inequality for open systems such that it gives the bound in terms of the time derivative of the eigenvalues of the density operator  rather then the time derivative of the density operator itself given by $\rho= \sum_{j}p_{j}\left|\psi_{j}\right\rangle \left\langle \psi_{j}\right|$ where $p_{j}$ are the eigenvalues and $\left|\psi_{j}\right\rangle $ are the eigenvectors of the density operator.  Start from 
\begin{equation}
\label{eq:1.1}
\sigma_{A}^{2}= \langle (\Delta A)^{2}\rangle=\sum_{j} p_{j}\langle \psi_{j} \vert A^{2}\vert \psi_{j}\rangle - \left( \sum_{j} p_{j}\langle \psi_{j} \vert A\vert \psi_{j}\rangle \right) ^{2},
\end{equation}
where $A$ is an observable and a Hamiltonian $H$ evolves the orthonormal basis of the density operator unitarily.  Then we can write  $\Delta A \stackrel{\text{def}}{=} A-\left\langle A\right\rangle $ which gives $\sigma^{2}_{A}=\left\langle A^{2}\right\rangle -\left\langle A\right\rangle ^{2}$.
Taking the time derivative of $\sigma_{A}$ yields
\begin{equation}
\label{eq:1.2}
\frac{\mathrm{d}}{\mathrm{d}t}\left( \sigma_{A}\right) =\frac{1}{2\sigma_{A}}\left[ \frac{\mathrm{d}\left\langle A^{2}\right\rangle }{\mathrm{d}t}-2\left\langle A\right\rangle \frac{\mathrm{d}\left\langle A\right\rangle }{\mathrm{d}t}\right]. 
\end{equation}
We can find the time derivatives of each term on the RHS of Eq. (\ref{eq:1.2}) as
\begin{equation}
\label{eq:1.3} 
\frac{\mathrm{d}\left\langle A\right\rangle }{\mathrm{d}t}=\sum_{j}\dot{p}_{j}\left\langle \psi_{j}\vert A\vert \psi_{j}\right\rangle +\left\langle \partial_{t}A\right\rangle +i\sum_{j}p_{j}\left\langle \psi_{j} \vert \left[ H,A\right] \vert \psi_{j} \right\rangle  .
\end{equation}
and
\begin{equation}
\label{eq:1.4} 
\frac{\mathrm{d}\left\langle A^{2}\right\rangle }{\mathrm{d}t}=\sum_{j}\dot{p}_{j}\left\langle \psi_{j}\vert A^{2}\vert \psi_{j}\right\rangle +\left\langle \partial_{t} A^{2}\right\rangle +i\sum_{j}p_{j}\left\langle \psi_{j} \vert \left[ H,A^{2}\right] \vert \psi_{j} \right\rangle ,
\end{equation}
where $\dot{p}_{j} \stackrel{\text{def}}{=}\mathrm{d}p_{j}/\mathrm{d}t$,  , and $p_{j} \geq 0$ are the eigenvalues of the density operator $\rho$ such that $\sum_{j}p_{j}=1$, yielding $\sum_{j}\dot{p}_{j}=0$.  We also used  $\frac{\mathrm{d}\langle A\rangle}{\mathrm{d}t}= \sum_{j}\frac{\mathrm{d}p_{j}}{\mathrm{d}t}\langle \psi_{j} \vert A \vert \psi_{j} \rangle + \left\langle  \partial_{t} A \right\rangle  + i\langle \left[ H,A\right]  \rangle$ to write Eq. (\ref{eq:1.4}) \cite{BAC}.  The results from Eqs. ( \ref{eq:1.2}- \ref{eq:1.4}) together with $\frac{\mathrm{d}(\sigma^{2}_{A})}{\mathrm{d}t}=\frac{\mathrm{d} \left\langle (\Delta A)^{2}\right\rangle }{\mathrm{d}t}$ yields \cite{BAC}
\begin{equation}
\label{eq:1}
2\sigma_{A}\dot{\sigma}_{A}=\sum_{j}\dot{p}_{j}\left\langle \psi_{j} \left|A^{2}\right|\psi_{j}\right\rangle -2\left\langle A\right\rangle \sum_{j}\dot{p}_{j}\left\langle \psi_{j}\left|A\right|\psi_{j}\right\rangle +i\left\langle \left[ H,A^{2}\right] \right\rangle -2i\left\langle A\right\rangle \left\langle \left[ H,A\right] \right\rangle +\left\langle \partial_{t}A^{2}\right\rangle  -2\left\langle A\right\rangle \left\langle \partial_{t}A\right\rangle .
\end{equation}
Knowing that
\begin{equation}
\label{eq:2}
\left\langle \psi_{j}\left|(\Delta A)^{2}\right|\psi_{j}\right\rangle = \left\langle \psi_{j}\left|A^{2}\right|\psi_{j}\right\rangle -2\left\langle A\right\rangle \left\langle \psi_{j}\left| A\right|\psi_{j}\right\rangle + \left\langle A\right\rangle ^{2} ,
\end{equation}
and multiplying both sides of Eq. (\ref{eq:2}) by $\dot{p}_{j}$ and summing over $j$ gives
\begin{equation}
\label{eq:3}
\sum_{j}\dot{p}_{j}\left\langle \psi_{j}\left|(\Delta A)^{2}\right|\psi_{j}\right\rangle = \sum_{j}\dot{p}_{j}\left\langle \psi_{j}\left|A^{2}\right|\psi_{j}\right\rangle -2\left\langle A\right\rangle \sum_{j}\dot{p}_{j}\left\langle \psi_{j}\left| A\right|\psi_{j}\right\rangle + \left\langle A\right\rangle ^{2} \sum_{j}\dot{p}_{j},
\end{equation}
where the last term is zero since $\Sigma_{j}\dot{p}_{j}=0$.  Next, use
\begin{equation}
\label{eq:4}
\left[H, A^{2}\right] =A\left[ H,A\right] +\left[ H, A\right] A=\left\lbrace A,\left[ H,A\right] \right\rbrace , 
\end{equation}
together with $\Delta A = A-\left\langle A\right\rangle $ to write
\begin{equation}
\label{eq:5}
\left\lbrace A,\left[ H, A\right] \right\rbrace =\left\lbrace \Delta A+\left\langle A\right\rangle I,\left[ H,A\right] \right\rbrace =\left\lbrace \Delta A,\left[ H,A\right] \right\rbrace +\left\lbrace \left\langle A\right\rangle I,\left[ H,A\right] \right\rbrace = \left\lbrace \Delta A,\left[ H,A\right] \right\rbrace + 2\left\langle A\right\rangle \left[ H,A\right] ,
\end{equation}
where $I$ is the identity matrix.  Taking the average value of Eq. (\ref{eq:5}) and rearranging give
\begin{equation}
\label{eq:6}
\left\langle \left\lbrace A,\left[ H,A\right] \right\rbrace \right\rangle -2\left\langle A\right\rangle \left\langle \left[ H,A\right] \right\rangle =\left\langle \left\lbrace \Delta A,\left[ H,A\right] \right\rbrace \right\rangle . 
\end{equation}
Finally, observe that
\begin{equation}
\begin{aligned}
\left\langle \left\lbrace \Delta A, \partial_{t}A\right\rbrace \right\rangle  &= \left\langle \left\lbrace \left( A-\left\langle A\right\rangle \right) , \partial_{t}A\right\rbrace \right\rangle  = \left\langle \left( A-\left\langle A\right\rangle \right) \partial_{t}A+\partial_{t}A\left( A-\left\langle A\right\rangle \right) \right\rangle \\
 &= \left\langle A\partial_{t}A\right\rangle -\left\langle A\right\rangle \left\langle \partial_{t}A\right\rangle +\left\langle \left( \partial_{t}A\right) A\right\rangle -\left\langle \partial_{t}A\right\rangle \left\langle A\right\rangle \\
  &= \left\langle \partial_{t}A^{2}\right\rangle -2\left\langle A\right\rangle \left\langle \partial_{t}A\right\rangle .
\end{aligned}
\label{eq:7}
\end{equation}
Using the results found in Eqs. (\ref{eq:5}-\ref{eq:7}), one can rewrite Eq. (\ref{eq:1}) as
\begin{equation}
\begin{aligned}
2\sigma_{A}\dot{\sigma}_{A} &= \sum_{j}\dot{p}_{j}\left\langle \psi_{j}\left|(\Delta A)^{2}\right|\psi_{j}\right\rangle  + i\left\langle \left\lbrace \Delta A, \left[ H,A\right] \right\rbrace \right\rangle + \left\langle \left\lbrace \Delta A,\partial_{t}A\right\rbrace \right\rangle \\
&= \sum_{j}\dot{p}_{j}\left\langle (\Delta A)^{2}\right\rangle_{j} + \left\langle \left\lbrace \Delta A, i\left[ H,A\right] +\partial_{t}A\right\rbrace \right\rangle .
\end{aligned}
\label{eq:8}
\end{equation}
Next letting $\frac{\mathrm{d}A}{\mathrm{d}t}=\dot{A}=i\left[ H,A\right] +\partial_{t}A$ and rewriting Eq. (\ref{eq:8}) yields
\begin{equation}
\label{eq:9}
2\sigma_{A}\dot{\sigma}_{A} = \sum_{j} \dot{p}_{j}\left\langle (\Delta A)^{2}\right\rangle _{j} + \left\langle \left\lbrace \Delta A, \dot{A} \right\rbrace \right\rangle = \sum_{j} \dot{p}_{j}\left\langle (\Delta A)^{2}\right\rangle _{j} + 2\,\cov\left( A, \dot{A}\right),
\end{equation}
where we have defined the covariance between two operators as 
$\cov(A,B)\defn \half \avg{A B + B A} - \avg{A}\,\avg{B}$.
Eq. (\ref{eq:9}) connects the time evolution of an observable's uncertainty to the dynamics of the quantum state.  In the next two sections we will study Eq. (\ref{eq:9}) further and investigate its connections with the Fisher information and quantum operators.

The result found in Eq. (\ref{eq:9}) can  be manipulated using the Cauchy-Schwarz inequality to give us an upper bound for fluctuation growth. 
Using the easily proved property that $2\, \cov(A,B) = \avg{\DA\, \DB} + \avg{\DB\,\DA}$ one then has upon invoking the Cauchy-Schwarz inequality
\be{GC5}
2 |\cov(A,B)| \le \sqrt{\avg{(\DA)^2} \,\avg{(\DB)^2}} +  \sqrt{\avg{(\DB)^2} \,\avg{(\DA)^2}}
= 2\,\sigma_A\,\sigma_B.
\ee
 Using \Eq{GC5} in \Eq{eq:9} allows us to write the following inequality
 \be{our:inequality}
 2 \sigma_A \dot{\sigma}_A \le \sum_j \dot{p}_j \bra{\psi_j} (\DA)^2 \ket{\psi_j} 
 +  2\,\sigma_A\,\sigma_{\dot{A}}.
 \ee
Note that \Eq{our:inequality} differs from our previous inequality in \Eq{eq:I1}, which upon taking a square root,  and using $(a + b)^2 \le 2 (a^2 + b^2)$ in reverse, takes the form
\be{our:previous:inequality}
2 \sigma_A \dot{\sigma}_A \le
\Tr[\dot{\rho} (\DA)^2] + 2\,\sigma_A\, \sigma_{(\pd_t A)}.
\ee
\Eq{our:previous:inequality} emphasizes the role of the full evolution of the density matrix with the explicit appearance of $\dot{\rho}$. In addition to the variance $\sigma_A$ of $A$ in the second term, it also involves  the variance $\sigma_{(\pd_t A)}$ of the explicit time dependence $\pd_t A$ of $A$.  
On the other hand, \Eq{our:inequality} emphasizes the role of the incoherent portion of the state evolution, i.e. the non-unitary portion of the evolution which changes the eigenvalues via $\dot{p}_j$. 
The the second term in \Eq{our:inequality}  now involves 
the variance  $\sigma_{\dot{A}}$ of the full time derivative 
$\dot{A} \defn \frac{d A}{dt} =\frac{\pd A}{\pd t} +  i\, [H, A]$
of $A$ coming from the coherent part of the evolution arising from the Hamiltonian, which cannot change the eigenvalues $p_i$, but instead merely rotates the basis vectors in Hilbert space.

In the following sections, we  will focus on Eq. (\ref{eq:9}) and view it in two different ways by  studying its connections with the Fisher information as well as the quantum operators.

\section{Equation (\ref{eq:9}) and the Fisher Information}

Quantum Fisher information (QFI) is a measure of how distinguishable the state is under a parameter change.    So, finding the relation between Eq. (\ref{eq:9}) and QFI can tell us how rapidly information about a parameter affects the fluctuations of an observable.  In this section we will focus on Eq. (\ref{eq:9}) and investigate its connection to the Fisher information.  

The Classical Fisher information (FI) quantifies how distinguishable nearby probability distributions are, such that large FI means that a small parameter change produces a noticeable statistical change \cite{FI1, FI2}.  Further, FI that is given by 
\begin{equation}
\label{eq:F1}
\mathcal{F}(\theta)=\sum_{x}\frac{1}{p(x|\theta)}\left( \frac{\partial p(x|\theta)}{\partial \theta}\right) ^{2},
\end{equation}
for a discrete random variable $\theta$ \cite{FI1}, is also a metric  on the space of probability distributions \cite{Rao}.  In a quantum system where a general evolution can be made by any arbitrary Hermitian operator, a  measure of distinguishability between two quantum states is given by the Quantum Fisher Information (QFI) 
\begin{equation}
\label{eq:F2}
\mathcal{F}_{Q}(\theta)=\sum_{k}\frac{(\partial_{\theta}p_{k})^{2}}{p_{k}}+2\sum_{k,l}\frac{(p_{k}-p_{l})^{2}}{p_{k}+p_{l}}\left|\left\langle \psi_{k}|\partial_{\theta}\psi_{l}\right\rangle \right|^{2} = \mathcal{F}_{I}+\mathcal{F}_{C}
\end{equation}
where the first term on the RHS is the contributions due to changes in eigenvalues and the second term on the RHS is due to changes in eigenvectors \cite{BC}.

To find the relation between Eq. (\ref{eq:9}) and the quantum Fisher information, we can rewrite Eq. (\ref{eq:9}) as
\begin{equation}
\label{eq:b1}
2\sigma_{A}\dot{\sigma}_{A} 
= Tr\left( \dot{\rho} (\Delta A)^{2}\right) -i \left\langle \left[ H,(\Delta A)^{2}\right] \right\rangle +\left\langle \left\lbrace \Delta A,\dot{A}\right\rbrace \right\rangle = Tr\left( \dot{\rho} (\Delta A)^{2}\right)+ \left\langle \left\lbrace \Delta A,\partial_{t}A\right\rbrace \right\rangle .
\end{equation}
A detailed derivation of Eq. (\ref{eq:b1}) is given in Appendix A.

Next, to find the relation between Eq. (\ref{eq:b1}) and the Fisher information, we use the definition of the symmetric logarithmic derivative (SLD) $L$ which generically defines the evolution of the density matrix via
\begin{equation}
\label{eq:b4}
\dot{\rho}\defn\frac{1}{2}\left\lbrace \rho,L\right\rbrace 
\end{equation}
Following the work of Garc{\'\i}a-Pintos, \tit{et al.} (2022)\cite{Gor},
we proceed by breaking  up both $\dot{\rho}$ and $L$ into their \tit{incoherent} and \tit{coherent} parts defined as their diagonal and off-diagonal components, respectively. For a general operator $A$ we define
\be{A:I:C}
A \equiv A_I + A_C, \quad\trm{where}\quad
A_I \defn \sum_{i} A_{i i} \ket{i}\bra{i}, \quad
A_C \defn \sum_{i\ne j}  A_{i j} \ket{i}\bra{j},
\ee
where $\ket{i}$ are the eigenstates of $\rho$.
We therefore write
\begin{equation}
\label{eq:b5}
\dot{\rho}=\dot{\rho}_{I}+\dot{\rho}_{C} \qquad L=L_{I}+L_{C},
\end{equation}
for the incoherent and coherent parts of $\rho$ and $L$.  

The motivation for dividing an operator into its diagonal and off-diagonal components stem from the 
general time evolution of the density matrix given as \cite{Gor}
\be{G6}
\dot{\rho} = i [\rho, H] + U(t)\,\dot{\chi}(t)\,U^\dag(t)  = i [\rho, H] + \sum_i \dot{p}_i(t) \ket{\psi_i(t)}\bra{\psi_i(t)} \equiv  i [\rho, H] +\sum_i \dot{p}_i \ket{i}\bra{i} .
\ee
Here, $\chi(t) \defn \sum_i \dot{p}_i(t) \ket{\psi_i(0)}\bra{\psi_i(0)} \equiv \sum_i \dot{p}_i(t) \ket{i_0}\bra{i_0}$,
is the diagonal matrix with the derivatives $\dot{p}_i(t)$ of the eigenvalues of $\rho(t)$, but with the initial
eigenstate $\ket{i_0} = \ket{\psi_i(0)}$.
The operator $U(t)$ is defined as the the solution to the Schr\"odinger equation such that $U(t)\ket{i_0} = \ket{i(t)} \defn \ket{i}$.
Taking matrix elements of \Eq{G6} yields 
\be{me:pi}
 \EV{i}{\dot{\rho}}{j} = i (p_i - p_j)\,H_{ij} + \dot{p}_i \d_{ij}, 
 %
 \implies
 (\dot{\rho})_I = \sum_i \dot{p}_i \ket{i}\bra{i} \quad\trm{and}\quad
 (\dot{\rho})_C = i \sum_{i\ne j} (p_i - p_j)\,H_{i j} \ket{i}\bra{j}.
\ee
Thus, we see that the coherent part $(\dot{\rho})_C$ of the state evolution is driven by the Hamiltonian for which the unitary evolution does not change the eigenvalues $p_i$. Conversely, the incoherent part of the state evolution $ (\dot{\rho})_I$ resulting from  non-unitary evolution changes the eigenvalues viz $\dot{p}_i$.
It is then easy to show that (see Appendix C)
\bea{}
\Tr[\dot{\rho} A]  = \Tr[( \dot{\rho})_I +\dot{\rho})_I) (A_I + A_C)] &=& \Tr[(\dot{\rho})_I A_I] + \Tr[(\dot{\rho})_C A_C], \no
\trm{with}\quad 
\Tr[(\dot{\rho})_I A_I] = \sum_i \dot{p}_i A_i &\;\trm{and}\;&
\Tr[(\dot{\rho})_C A_C]  = i \,\Tr\big[ [\rho, H] A_C\big]
\eea

We can then write Eq. (\ref{eq:b4}) as
\begin{equation}
\label{eq:b6}
\dot{\rho}_{I}+\dot{\rho}_{C}=\thalf\left\lbrace \rho,L_{I}+L_{C}\right\rbrace, 
\end{equation}
which gives
\begin{equation}
\label{eq:b7}
\dot{\rho}_{I}=\thalf\left\lbrace \rho,L_{I}\right\rbrace = \thalf \{(\rho)_I L_I\}
 \qquad  \dot{\rho}_{C}=\thalf\left\lbrace \rho,L_{C}\right\rbrace = \thalf \{(\rho)_C L_C\},
\end{equation}
where the incoherent and coherent parts of $L$ are defined as \cite{Gor}
\be{LI:LC}
L_I = \sum_i \frac{d\ln p_i}{dt} \ket{i}\bra{i}, \quad\trm{and}\quad
L_C = 2\,i\, \sum_{i\ne j} \frac{\bra{i} [\rho, H]\ket{j}}{(p_i + p_j)} \ket{i}\bra{j}.
\ee

Using Eq. (\ref{eq:b5}) we can write
\begin{equation}
\label{eq:b8}
\Tr\left( \dot{\rho}(\Delta A)^{2}\right) =\Tr\left( \dot{\rho}_{I}(\Delta A)^{2}\right) + \Tr\left( \dot{\rho}_{C}(\Delta A)^{2}\right),
\end{equation}
where we can rewrite the first term using Eq. (\ref{eq:b7}), as
\begin{equation}
\label{eq:b9}
\Tr\left( \dot{\rho}_{I}(\Delta A)^{2}\right) = \Tr\left( \thalf\left\lbrace \rho,L_{I}\right\rbrace (\Delta A)^{2}\right) = \thalf\Tr\left( \rho\left\lbrace L_{I},(\Delta A)^{2}\right\rbrace \right)  = \thalf\left\langle \left\lbrace L_{I},(\Delta A)^{2}\right\rbrace  \right\rangle .
\end{equation}
Recalling that 
\begin{equation}
\label{eq:b10}
\cov(L_{I},(\Delta A)^{2})=\thalf\left\langle \left\lbrace L_{I},(\Delta A)^{2}\right\rbrace \right\rangle-\left\langle L_{I} \right\rangle \left\langle (\Delta A)^{2}\right\rangle ,
\end{equation}
and utilizing 
\begin{equation}
\label{eq:b11}
\Tr\left( \dot{\rho}_{I}\right)=\thalf \Tr\left( \rho L_{I}+L_{I}\rho\right)  =\Tr\left( \rho L_{I}\right) = \left\langle L_{I}\right\rangle = \sum_i \dot{p}_i = 0,
\end{equation}
Eq. (\ref{eq:b9}) yields
\begin{equation}
\label{eq:b12}
\Tr\left( \dot{\rho}_{I}(\Delta A)^{2}\right)=\cov(L_{I},(\Delta A)^{2})=\cov(L_{I},((\Delta A)^{2})_{I}).
\end{equation}
In the above, the last equality is due to the fact that $\rho$ is diagonal in its eigenbasis, and $L_{I}$ is diagonal, so only the diagonal part of $(\Delta A)^{2}$ contributes.  More precisely, $\cov\left( L_{I},(\Delta A)^{2}\right) =\cov\left( L_{I},((\Delta A)^{2})_{I}\right)+\cov\left( L_{I},((\Delta A)^{2})_{C}\right)$ where $\cov\left( L_{I},((\Delta A)^{2})_{C}\right)=\frac{1}{2}Tr\left( \rho\left\lbrace L_{I},((\Delta A)^{2})_{C}\right\rbrace \right) -\left\langle L_{I}\right\rangle \left\langle ((\Delta A)^{2})_{C}\right\rangle =0$ in which all terms are zero.  Similarly, 
\begin{equation}
\label{eq:b13}
Tr\left( \dot{\rho}_{C}(\Delta A)^{2}\right)=\cov(L_{C},((\Delta A)^{2})_{C}).
\end{equation}
We can also write
\begin{equation}
\label{eq:b14}
\left\langle \left\lbrace \Delta A, \partial_{t}A\right\rbrace \right\rangle = 2\,\cov\left( \Delta A, \partial_{t}A\right) = 2\,\cov\left( A-\left\langle A\right\rangle ,\partial_{t}A\right) =2\,\cov\left( A, \partial_{t}A\right) .
\end{equation}
Substituting the results found in Eqs. (\ref{eq:b12},\ref{eq:b13}, \ref{eq:b14}) into Eq. (\ref{eq:b1}) yields
our reformulated equality in \Eq{eq:9} as
\begin{equation}
\label{eq:b15}
2\sigma_{A}\dot{\sigma}_{A}=\cov\left( L_{I},((\Delta A)^{2})_{I}\right) +\cov\left( L_{C},((\Delta A)^{2})_{C}\right)+2\,\cov\left( A,\partial_{t}A\right) .
\end{equation}

Knowing that $\mathcal{F}_{I}=\left\langle \left( L_{I}\right) ^{2}\right\rangle = \sigma^2_{L_I}$ 
and $\mathcal{F}_{C}=\left\langle \left( L_{C}\right) ^{2}\right\rangle = \sigma^2_{L_C}$, 
and using the Cauchy-Schwarz inequality we have
\begin{equation}
\label{eq:b16}
\cov\left( L_{I},((\Delta A)^{2})_{I}\right) \leq \sqrt{\mathcal{F}_{I}}\: \sigma_{((\Delta A)^{2})_{I}} \qquad  \cov\left( L_{C},((\Delta A)^{2})_{C}\right) \leq \sqrt{\mathcal{F}_{C}}\: \sigma_{((\Delta A)^{2})_{C}},
\end{equation}
where $\sigma^{2}_{((\Delta A)^{2})}= \left\langle (\Delta A)^{4}\right\rangle -\left\langle (\Delta A)^{2}\right\rangle ^{2}$.  We can also use the Cauchy-Schwarz to rewrite $\cov\left( A,\partial_{t}A\right) $ as
\begin{equation}
\label{eq:b17}
\cov\left( A,\partial_{t}A\right)  \leq \sigma_{A} \sigma_{\partial_{t}A}.
\end{equation}
Using Eqs. (\ref{eq:b16},\ref{eq:b17}), we can rewrite the inequality associated with Eq. (\ref{eq:b15}) as
\begin{equation}
\label{eq:b18}
2\sigma_{A}\dot{\sigma}_{A} \leq \sqrt{\mathcal{F}_{I}}\: \sigma_{((\Delta A)^{2})_{I}} + \sqrt{\mathcal{F}_{C}}\: \sigma_{((\Delta A)^{2})_{C}} + 2 \sigma_{A} \sigma_{\partial_{t}A}.
\end{equation} 

In summary, since
$(\dot{\rho})_I=\sum_i\dot{p}_i \ket{i}\bra{i}$ is diagonal, and 
 $(\dot{\rho})_C = i [\rho, H] = i \sum_{j\ne k}[\rho,H]_{jk} \ket{j}\bra{k}$, is off-diagonal 
 we have for \tit{any} operator  operator $\mO\in\{A,A^2,\ldots\}$ that
\begin{align}
\frac{d}{dt}\avg{\mO} 
&= \Tr[\dot{\rho} \mO] + \avg{\dot{\mO}}= 
\Tr[\big( (\dot{\rho})_I+ (\dot{\rho})_C\big) \big(\mO_I + \mO_C\big)] + \avg{\dot{\mO}} \no
&=  
 \Tr[(\dot{\rho})_I \mO_I] + \Tr[(\dot{\rho})_C \mO_C] + \avg{\dot{\mO}}, \no
 &\equiv \trm{cov}(L_I, \mO_I ) +  \trm{cov}(L_C, \mO_C )+ \avg{\dot{\mO}}, \\
 \implies 
 \left|\frac{d}{dt}\avg{\mO}\right| &\le \sqrt{F_I}\,\sigma_{\mO_I} + \sqrt{F_C}\,\sigma_{\mO_C} + |\avg{\dot{\mO}}|
\end{align}
 
Eq. (\ref{eq:b18}), which uses $\mO = (\DA)^2$, demonstrates variations of uncertainty of observable $A$ with the coherent and incoherent Fisher information, showing that  observable fluctuations cannot evolve faster than allowed by the information geometry of the state.  The first term on the RHS shows the contributions of the incoherent part of the dynamics, which can be the changes due to the environment, while the second term on the RHS captures the contributions from coherent evolutions which is due to the unitary evolution.  The last term on the RHS is due to the changes of the observable itself over time.  Thus this equation can tell us where the change in the uncertainty of the observable is coming from.

For an observable $A$, the quantity $\operatorname{Var}[(\Delta A)^2]=\langle(\Delta A)^4\rangle-\langle(\Delta A)^2\rangle^2$ characterizes the fourth-order statistics of measurement outcomes and can complement established measures of quantum uncertainty, quantum-state distinguishability, and dynamical evolution.  It is directly related to the fourth central moment and the kurtosis of the measurement distribution, since kurtosis is defined as the standardized fourth central moment given by $\kappa =\frac{\left\langle (\Delta A)^4\right\rangle }{\left\langle (\Delta A)^{2}\right\rangle ^{2}}=\frac{\mu_4}{(\sigma_A)^4}$ \cite{West}. 
In mathematical statistics, skewness and kurtosis are two quantities used to characterize the shape of a probability distribution. Skewness involves the third moment of the distribution and is a measure of the “asymmetry” of the probability distribution, and gives the leading order deviation from a Gaussian.
Kurtosis, on the other hand, involves the fourth moment of the distribution and
is a measure of the “tailedness” of the probability distribution (i.e., a quantifier of how often outliers occur).
Recently, the kurtosis has found applicability to the curvature and torsion of quantum mechanical evolutions in both stationary and non-stationary Hamiltonians in terms of a Frenet-Serret apparatus
\cite{Alsing_Cafaro_FSA:I:2024, Alsing_Cafaro_FSA:II:2024}.
Its time derivative, $\frac{d}{dt}\operatorname{Var}[(\Delta A)^2]$, therefore describes how the distribution of fluctuation magnitudes changes as the quantum state evolves, potentially revealing dynamical features that are not captured by the mean or variance alone.  This perspective is particularly relevant to open quantum systems, in which interactions with the environment can modify quantum state coherence and statistical properties of observables \cite{BP}, and to quantum metrology, where measurement statistics and quantum noise constrain the precision with which physical parameters can be estimated \cite{Gi} where coherent evolution and environmental interactions can modify observable statistics, and to quantum metrology, where fluctuations constrain the precision with which physical parameters can be estimated.  Research on quantum speed limits has established that the rate of evolution of observables is constrained by statistical fluctuations and information-theoretic quantities, including quantum Fisher information \cite{DC}\cite{CEP}\cite{Gor}. Motivated by these results, the time derivative of $\operatorname{Var}[(\Delta A)^2]$ can be investigated as a complementary indicator of the dynamical restructuring of quantum fluctuations, with potential applications to the characterization of non-Gaussian statistics, decoherence, and collective quantum systems. Its precise operational significance and relationship to established information-theoretic measures, however, require separate derivation and investigation.

Next, we will focus on viewing Eq. (\ref{eq:9}) in terms of quantum operations \cite{NCh}.  In particular, we will express the time derivative of the eigenvalues $\dot{p}_{i}(t)$ in terms of the Kraus operators.  Then using the relation between the Kraus and the Lindblad operators, $\dot{p}_{i}(t)$ can be expressed in terms of Lindblad operators as well.

\section{Evolution of the Density Operator's Eigenvalues \\ in Terms of  Kraus operators}

To find the connection between the measurable quantity $p_{j}$ and the dynamical description of the quantum process that is expressed by Kraus or Lindblad operators, we will find the relation between $\dot{p}_{j}$ and Kraus operators in this section, while we will investigate the relations between $\dot{p}_{j}$ and Lindblad operators in section V.

Kraus operator $E_{k}$ is one of a set of $\left\lbrace E_{k}\right\rbrace $ that describes the most general evolution of a density matrix under a quantum channel \cite{NCh}.  They are simply a mathematical tool to describe how a quantum system changes when it interacts with its environment.  Considering each possible interaction as a Kraus operator, the most general evolution of a density matrix is written as
\begin{equation}
\label{eq:15}
\rho(t+\mathrm{d}t)=\sum_{k} E_{k}(\mathrm{d}t)\rho(t) E^{\dagger}_{k}(\mathrm{d}t)
\end{equation} 
with
\begin{equation}
\label{eq:16}
\sum_{k} E^{\dagger}E_{k}=1
\end{equation}
where $E_{k}(\mathrm{d}t)$ are Kraus operators describing the system-environment interaction during an \textit{arbitrary} discrete time step $\mathrm{d}t$ (not necessarily small) with the condition given by Eq. (\ref{eq:16}) to ensure a completely positive and trace preserving (CPTP) map \cite{NCh}.

Using $\rho=\sum_{j}p_{j}\left|\psi_{j}\right\rangle \left\langle \psi_{j}\right|$ Eq. (\ref{eq:15}) can be rewritten as
\begin{equation}
\label{eq:17}
\rho(t+\mathrm{d}t)=\sum_{kj}p_{j}E_{k}\left|\psi_{j}\right\rangle \left\langle \psi_{j}\right|E^{\dagger}_{k}, 
\end{equation} 
where we have set $E_{k}(\mathrm{d}t)=E_{k}$ and $E^{\dagger}_{k}(\mathrm{d}t)=E^{\dagger}_{k}$ for simplicity.  Knowing that $\left|\psi_{j}\right\rangle $ are eigenvectors of $\rho$ and $p_{j}$ are its eigenvalues, the population of eigenstate $\left|\psi_{i}\right\rangle $ at time $t+\mathrm{d}t$ is given by 
\begin{equation}
\label{eq:18}
p_{i}(t+\mathrm{d}t)=\left\langle \psi_{i}|\rho(t+\mathrm{d}t)|\psi_{i}\right\rangle .
\end{equation}
Allowing now that the time step $dt$ can be small,  the time derivative of $p_{i}$ can be written as
\begin{equation}
\label{eq:19}
\dot{p}_{i}(t)=\lim_{\mathrm{d}t \to 0}\frac{p_{i}(t+\mathrm{d}t)-p_{i}(t)}{\mathrm{d}t},
\end{equation}
which can be rewritten in terms of the corresponding density operator given by Eq. (\ref{eq:18}) as
\begin{equation}
\label{eq:20}
\dot{p}_{i}(t)=\lim_{\mathrm{d}t \to 0}\frac{\left\langle \psi_{i}|\rho(t+\mathrm{d}t)|\psi_{i}\right\rangle -p_{i}(t)}{\mathrm{d}t}.
\end{equation}
Substituting $\rho(t+\mathrm{d}t)$ from Eq. (\ref{eq:15}) yields
\begin{equation}
\label{eq:21}
\dot{p}_{i}(t)=\lim_{\mathrm{d}t \to 0}\frac{\sum_{k}\left\langle \psi_{i}|E_{k}\rho(t)E^{\dagger}_{k}|\psi_{i}\right\rangle -p_{i}(t)}{\mathrm{d}t}.
\end{equation}
The result found in Eq. (\ref{eq:21}) is the evolution of the eigenvalue of the density operator with respect to time expressed in terms of the Kraus operators.  We can now consider a simple example to determine $\dot{p}_{i}$ for a two-level system using Eq. (\ref{eq:21}).
 
\subsection{Example of Eq. (\ref{eq:21}): Amplitude Damping using the Kraus Operators} 

Suppose a two-level system has an evolution $\Omega=\omega/2 \sigma_{z}$ and a decay process with rate $\gamma$.  If state $\left|0\right\rangle $ with probability $p_{0}$, and state $\left|1\right\rangle$ with probability $p_{1}$ indicate the ground state and the excited state and their associated probabilities, respectively, and the Kraus operators for amplitude damping are given by
\begin{equation}
\label{eq:22}
E_{0}=\begin{pmatrix}
1 & 0\\
0 & \sqrt{1-\gamma \mathrm{d}t}
\end{pmatrix}
\quad\text{and}\quad
E_{1}=\begin{pmatrix}
0 & \sqrt{\gamma \mathrm{d}t} \\ 
0 & 0
\end{pmatrix}
\end{equation} 
where $\gamma$ is the decay rate, then the requirement for Kraus operators given by Eq. (\ref{eq:16}) for this system yields 
\begin{equation}
\label{eq:23}
E^{\dagger}_{0}E_{0}+E^{\dagger}_{1}E_{1}=I ,
\end{equation}
where $I$ is the identity matrix.  Assume $p_{0}+p_{1}=1$ and the density matrix is given by 
\begin{equation}
\label{eq:24}
\begin{pmatrix}
p_{0} & 0 \\ 0 & p_{1}
\end{pmatrix},
\end{equation}
and exploit Eq. (\ref{eq:16}) to get
\begin{equation}
\label{eq:25}
\begin{pmatrix}
p_{0} & 0 \\ 0 & (1-\gamma \mathrm{d}t)p_{1} 
\end{pmatrix}
+
\begin{pmatrix}
\gamma p_{1} \mathrm{d}t  & 0 \\ 0 & 0
\end{pmatrix}
=
\begin{pmatrix}
p_{0}+\gamma p_{1} \mathrm{d}t  & 0 \\ 0 & (1-\gamma \mathrm{d}t)p_{1}
\end{pmatrix}.
\end{equation} 
Then using Eq. (\ref{eq:21}) to find $\dot{p}_{0}$ and $\dot{p}_{1}$, we obtain
\begin{equation}
\label{eq:26}
\dot{p}_{0}=\lim_{\mathrm{d}t \to 0} \frac{\left\langle 0\right| \sum_{k} E_{k}\rho E^{\dagger}_{k} \left| 0\right\rangle -p_{0}}{\mathrm{d}t}=\lim_{dt \to 0}\frac{p_{0}+\gamma \mathrm{d}t  p_{1}-p_{0}}{dt}=\gamma p_{1},
\end{equation} 
and
\begin{equation}
\label{eq:27}
\dot{p}_{1}=\lim_{dt \to 0} \frac{\left\langle 1\right| \sum_{k} E_{k}\rho E^{\dagger}_{k} \left| 1\right\rangle -p_{1}}{dt}=\lim_{dt \to 0}\frac{(1-\gamma dt)p_{1}-p_{1}}{dt}=-\gamma p_{1}.
\end{equation} 
As expected, the results found in Eqs. (\ref{eq:26}, \ref{eq:27}) indicate that the probability lost in the excited state  $\left|1\right\rangle $ is gained in the ground state $\left| 0\right\rangle $.  We will revisit amplitude damping in Section VI.A using the Lindblad operators and evaluate Eq. (\ref{eq:9}).

Next, we will use the relation between the Kraus and the Lindblad operators operators to express Eq. (\ref{eq:27}) in terms of Lindblad operators. 

\section{Evolution of the Density Operator's Eigenvalues in Terms of Lindblad Operators }

Alternatively, one can rewrite Kraus operators in terms of Lindblad operators for a Hamiltonian $H$ and find a relation for the time evolution of $p_{i}$ in terms of  Lindblad operators.  Using the time derivative of the density matrix as given in Ref. \cite{H}, we can write Kraus operators for infinitesimal time steps in terms of a Hamiltonian evolution $H$ as  
\begin{equation}
\begin{aligned}
E_{0} &=  I-iH \mathrm{d}t-\frac{1}{2}\sum_{k\geq 1}L^{\dagger}_{k}L_{k} \mathrm{d}t\\
E_{k} &= \sqrt{dt}L_{k},
\label{eq:28}
\end{aligned}
\end{equation}
where $L_{k}$ are Lindblad operators , $I$ is the identity matrix and $E_{k}$ indicate the dissipative terms, such that $E^{\dagger}_{0}E_{0}+\sum_{k}E^{\dagger}_{k}E_{k}=I$.  The density operator at time $t+\mathrm{d}t$ can be written as
\begin{equation}
\label{eq:29}
\rho(t+\mathrm{d}t)=E_{0} \rho (t) E^{\dagger}_{0} + \sum_{k\geq 1}E_{k}\rho (t) E^{\dagger}_{k}.
\end{equation}
Upon substituting Eq. (\ref{eq:28}) into Eq. (\ref{eq:29}), we find
\begin{equation}
\label{eq:29.1}
\rho(t + \mathrm{d}t)\simeq \rho(t) - \frac{i}{\hbar}\left[ H,\rho\right] \mathrm{d}t + \sum_{k \geq 1}\left( L_{k} \rho L^{\dagger}_{k} -\frac{1}{2}\left\lbrace L^{\dagger}_{k}K_{k},\rho\right\rbrace \right) \mathrm{d}t,
\end{equation}
which after a simple rearrangement  gives
\begin{equation}
\label{eq:29.2}
\lim_{\mathrm{d}t \to 0} \frac{\rho(t+\mathrm{d}t)-\rho(t)}{\mathrm{d}t}=\frac{\mathrm{d}\rho}{dt}\simeq -\frac{i}{\hbar} \left[ H, \rho\right] +\sum_{k\geq 1}\left( L_{k}\rho L^{\dagger}_{k}-\frac{1}{2}\left\lbrace L^{\dagger}_{k}L_{k}, \rho\right\rbrace \right) .
\end{equation}
Eq. (\ref{eq:29.2}) is the Lindblad master equation, which demonsrtates that Eq. (\ref{eq:28}) indeed give the relations between Lindblad and Kraus operators.

Next, use Eq. (\ref{eq:29}) to write Eq. (\ref{eq:18}) as
\begin{equation}
\label{eq:30}
p_{i}(t+\mathrm{d}t)=\left\langle \psi_{i}\right| \rho(t+\mathrm{d}t)\left|\psi_{i}\right\rangle =\sum_{k}\left\langle \psi_{i}\right|E_{k}\rho E^{\dagger}_{k}\left|\psi_{i}\right\rangle + \left\langle \psi_{i}\right|E_{0}\rho E^{\dagger}_{k}\left|\psi_{i}\right\rangle ,
\end{equation}
where we have set $\rho (t)=\rho$, for simplicity.  By substituting the terms from Eq. (\ref{eq:28}) into Eq. (\ref{eq:19}), we will obtain the time evolution of $p_{i}$ in terms of the Lindblad operators as 
\begin{equation}
\label{eq:37}
\dot{p}_{i} = \sum_{kj} \left( p_{j}\left|\left\langle \psi_{i}\right|L_{k}\left|\psi_{j}\right\rangle \right|^{2}-p_{i}\left|\left\langle \psi_{j}\right|L_{k}\left|\psi_{i}\right\rangle \right|^{2}\right) .
\end{equation}
Eq. (\ref{eq:37}) is the Lindblad representation of the time derivative of the density operator's eigenvalues.  A complete derivation of Eq. (\ref{eq:37}) is given in Appendix B.  Next, we will look at a few well-known  examples and compute $\dot{p}_{i}$ in each case using the corresponding Lindblad operators.

\section{Examples of Eq. (\ref{eq:37})}

In this section we will use Eq. (\ref{eq:37}) to solve some examples using the Lindblad operators corresponding the Kraus operators.

\subsection{Amplitude Damping}
We first consider the amplitude damping example from the  previous section with a two-level system and an evolution $\Omega=\omega/2 \sigma_{z}$ and a decay process with rate $\gamma$.  We considered state $\left|0\right\rangle $ with probability $p_{0}$, and state $\left|1\right\rangle$ with probability $p_{1}$ as the ground state and the excited state and their associated probabilities, respectively, and the Kraus operators for amplitude damping are given by
\begin{equation*}
E_{0}=\begin{pmatrix}
1 & 0\\
0 & \sqrt{1-\gamma \mathrm{d}t}
\end{pmatrix}
\quad\text{and}\quad
E_{1}=\begin{pmatrix}
0 & \sqrt{\gamma \mathrm{d}t} \\ 
0 & 0
\end{pmatrix}
\end{equation*} 
where $\gamma$ is the decay rate.  To find the rate of change of probabilities we will compute $\dot{p}_{i}$ using Eq. (\ref{eq:37}).  According to Eq. (\ref{eq:28}), one can write $L_{1}$ as $L_{1}=E_{1}/\sqrt{\mathrm{d}t}$ which gives
\begin{equation}
\label{eq:38}
L_{1}=\sqrt{\gamma}\left|0\right\rangle \left\langle 1\right|=\begin{pmatrix}
0 & \sqrt{\gamma}\\
0 & 0
\end{pmatrix}
\end{equation}
Using Eq. (\ref{eq:38}) one can compute $\dot{p}_{0}$ and $\dot{p}_{1}$ from Eq. (\ref{eq:37}) as 
\begin{equation}
\begin{aligned}
\dot{p}_{0} &= p_{1}\left|\left\langle 0\right|L_{1}\left|1\right\rangle \right|^{2}-p_{0}\left|\left\langle 1\right|L_{1}\left|0\right\rangle \right|^{2} = \gamma p_{1}\\
\dot{p}_{1} &= p_{0}\left|\left\langle 1\right|L_{1}\left|0\right\rangle \right|^{2}-p_{1}\left|\left\langle 0\right|L_{1}\left|1\right\rangle \right|^{2} = -\gamma p_{1}
\end{aligned}
\label{eq:39}
\end{equation}
where we have used
\begin{equation}
\begin{aligned}
\left\langle 0\right|L_{1}\left|0\right\rangle &= \sqrt{\gamma}\left\langle 0|0\right\rangle \left\langle 1|0\right\rangle =0\\
\left\langle 0\right|L_{1}\left|1\right\rangle &= \sqrt{\gamma}\left\langle 0|0\right\rangle \left\langle 1|1\right\rangle =\sqrt{\gamma}\\
\left\langle 1\right|L_{1}\left|0\right\rangle &= 0\\
\left\langle 1\right|L_{1}\left|1\right\rangle &= 0 .
\end{aligned}
\label{eq:40}
\end{equation}
The result found in Eq. (\ref{eq:39}) indicates that the probability flows from the first excited state to the ground state, which is the same result found in Eqs.(\ref{eq:26}, \ref{eq:27}) using the Kraus operators.

To verify Eq. (\ref{eq:9}), consider observable $A=J_{z}$, which gives
\begin{equation}
\begin{aligned}
\left\langle A\right\rangle &= Tr(\rho A)= \frac{1}{2}\left( p_{0}-p_{1}\right), \qquad \left\langle A^{2}\right\rangle =\frac{1}{4} \\
\sigma_{A} &= \sqrt{p_{0}p_{1}} ,
\end{aligned}
\label{eq:40.1}
\end{equation} 
where we have used   
\begin{equation}
\label{eq:40.1.1}
\rho=
\begin{pmatrix}
p_{0} & 0\\
0 & p_{1}
\end{pmatrix},
\end{equation}
with $p_{0}+p_{1}=1$.  We can use the results found in Eq. (\ref{eq:39}) to  obtain
\begin{equation}
\label{eq:40.2}
\dot{\sigma}_{A}=\frac{\dot{p}_{0}p_{1} + p_{0}\dot{p}_{1}}{2\sqrt{p_{0}p_{1}}}=\gamma p_{1}\left( p_{1}-p_{0}\right) .
\end{equation}
Note that the factor $2\,\sigma_A$ in the denominator in \Eq{eq:40.2} has factored in the final expression. 
Furthermore, using $\Delta A=A-\left\langle A\right\rangle  I$, we find
\begin{equation}
\begin{aligned}
\left\langle 0\right|(\Delta A)^{2}\left|0\right\rangle &=\left\langle (\Delta A)^{2}\right\rangle _{0} = p^2_{1} , \qquad  \left\langle 1\right|(\Delta A)^{2}\left|1\right\rangle =\left\langle (\Delta A)^{2}\right\rangle _{1}=p^2_{0} \\ 
\sum_{j}\dot{p}_{j}\left\langle (\Delta A)^{2}\right\rangle _{j} &= \dot{p}_{0}\left\langle (\Delta A)^{2}\right\rangle _{0} + \dot{p}_{1}\left\langle (\Delta A)^{2}\right\rangle _{1}=\dot{p}_{0} p^{2}_{1} + \dot{p}_{1} p^{2}_{0} = \gamma p_{1} \left( p_{1} - p_{0}\right) .
\end{aligned}
\label{eq:40.3}
\end{equation}
Comparing the results from Eqs. (\ref{eq:40.2},\ref{eq:40.3}), along with the fact that the observable has no explicit time derivative, we have $cov\left( A, \dot{A}\right) =0$, and we obtain 
\begin{equation}
\label{eq:40.4}
2 \sigma_{A}\dot{\sigma}_{A}= \sum_{j}\left\langle (\Delta A)^{2}\right\rangle_{j}=  \gamma p_{1} \left( p_{1} - p_{0}\right). 
\end{equation}
Eq. (\ref{eq:40.4}) indicates how the the variance in $A$ changes with time.  Since $\gamma , p_0,p_1 >0$,  the variance increases when $p_1>p_0$ and decreases when $p_0>p_1$, and it becomes zero when $p_1=p_0$ or when $p_1=0$.  Thus decay tends to reduce the variance when the lower level is more populated, whereas the variance can increase when the excited level is more populated, before the system approaches its steady state.  Furthermore, the coherent evolution generated by the Hamiltonian does not affect the populations, so the changes in the variance are governed by the dissipative process which is due to $A$.     

\subsection{Bit Flip}
Our second example is the bit flip channel where the probability of switching from $\left|0\right\rangle $ to $\left|1\right\rangle $ is $1-p$.  The Kraus operators corresponding to this process are given by \cite{NCh}
\begin{equation}
\label{eq:41}
E_{0}=\sqrt{p}\ I=\begin{pmatrix}
\sqrt{p} & 0\\
0 & \sqrt{p}
\end{pmatrix}
\quad\text{and}\quad
E_{1}=\sqrt{1-p}\ \sigma_{x}=\begin{pmatrix}
0 & \sqrt{1-p} \\ 
\sqrt{1-p} & 0
\end{pmatrix} .
\end{equation}  
To find the Lindblad operator using Eq. (\ref{eq:28}) we have $E_{1}=\sqrt{\mathrm{d}t}\ L_{1}$, which after comparing it with Eq. (\ref{eq:38}) gives
\begin{equation}
\label{eq:42}
L_{1}=\sqrt{\frac{1-p}{\mathrm{d}t}}\ \sigma_{x}.
\end{equation}
Assuming continuous time evolution, we can write
\begin{equation}
\label{eq:43}
\lim_{dt \to 0}\frac{1-p}{dt}=\gamma ,
\end{equation}
which can be rewritten as
\begin{equation}
\label{eq:44}
1-p=\gamma/ \mathrm{d}t.
\end{equation}
Using Eq. (\ref{eq:41}), the Lindblad operator can be written as
\begin{equation}
\label{eq:45}
L_{1}=\sqrt{\gamma}\ \sigma_{x}=\sqrt{\gamma}\left( \left|0\right\rangle \left\langle 1\right|+\left|1\right\rangle \left\langle 0\right|\right) .
\end{equation}
Using Eq. (\ref{eq:42}), we can now find $\dot{p}_{0}$ and $\dot{p}_{1}$ as
\begin{equation}
\begin{aligned}
\dot{p}_{0} &= p_{1}\left|\left\langle 0\right|L_{1}\left|1\right\rangle \right|^{2} - p_{0}\left|\left\langle 1\right|L_{1}\left|0\right\rangle \right|^{2}=\gamma (p_{1}-p_{0})\\
\dot{p}_{1} &= p_{0}\left|\left\langle 1\right|L_{1}\left|0\right\rangle \right|^{2} - p_{1}\left|\left\langle 0\right|L_{1}\left|1\right\rangle \right|^{2}=\gamma (p_{0}-p_{1}),
\end{aligned}
\label{eq:46}
\end{equation}
where we have used $\left\langle 0\left|L_{1}\right|0\right\rangle = \left\langle 1\left|L_{1}\right|1\right\rangle=0$ and $\left\langle 0\left|L_{1}\right|1\right\rangle = \left\langle 1\left|L_{1}\right|0\right\rangle = \sqrt{\gamma}$.  
  
This indicates a symmetric probability exchange between the ground state and the first excited state.  If $p_{1}>p_{0}$, then $\dot{p}_{0}> 0$ and $\dot{p}_{1}< 0$, so probability floes from the first excited state to the ground state.  We will verify Eq. (\ref{eq:9}) for the bit-phase flip example.

\subsection{Phase Flip}
In this example we consider the phase flip channel in which the Kraus operators are given as \cite{NCh}
\begin{equation}
\label{eq:47}
E_{0}=\sqrt{p}\ I=\begin{pmatrix}
\sqrt{p} & 0\\
0 & \sqrt{p}
\end{pmatrix}
\quad\text{and}\quad
E_{1}=\sqrt{1-p}\ \sigma_{z}=\begin{pmatrix}
\sqrt{1-p} & 0 \\ 
0 & -\sqrt{1-p}
\end{pmatrix}
\end{equation}
Exploiting the continuity condition for time evolution and Eq. (\ref{eq:41}), we can find the Lindblad operator corresponding to this process as
\begin{equation}
\label{eq:48}
L_{1}=\sqrt{\gamma}\ \sigma_{z}=\sqrt{\gamma}\left( \left|0\right\rangle \left\langle 0\right| - \left|1\right\rangle \left\langle 1\right|\right) .
\end{equation} 
Then using $\left\langle 0\right|L_{1}\left|0\right\rangle = \left\langle 1\right|L_{1}\left|1\right\rangle = \sqrt{\gamma}$ and $\left\langle 0\right|L_{1}\left|1\right\rangle = \left\langle 1\right|L_{1}\left|0\right\rangle = 0$, we find $\dot{p}_{0}$ and $\dot{p}_{1}$ as
\begin{equation}
\begin{aligned}
\dot{p}_{0} &= p_{1}\left|\left\langle 0\right|L_{1}\left|1\right\rangle \right|^{2} - p_{0}\left|\left\langle 1\right|L_{1}\left|0\right\rangle \right|^{2}=0\\
\dot{p}_{1} &= p_{0}\left|\left\langle 1\right|L_{1}\left|0\right\rangle \right|^{2} - p_{1}\left|\left\langle 0\right|L_{1}\left|1\right\rangle \right|^{2}=0.
\end{aligned}
\label{eq:49}
\end{equation}
The result found in Eq. (\ref{eq:49}) indicates that there is no probability exchange.  Rather, it is only the phase that is flipped.  So, the probabilities are fixed but information in phase is lost.

\subsection{Bit-Phase Flip}
In this example we would like to calculate the bit-phase flip for which the Kraus operators are given as \cite{NCh}
\begin{equation}
\label{eq:50}
E_{0}=\sqrt{p}\ I=\begin{pmatrix}
\sqrt{p} & 0\\
0 & \sqrt{p}
\end{pmatrix}
\quad\text{and}\quad
E_{1}=\sqrt{1-p}\ \sigma_{y}=\begin{pmatrix}
0 & -i\sqrt{1-p} \\ 
i\sqrt{1-p} & 0
\end{pmatrix}
\end{equation}
and the Lindblad operator is given by
\begin{equation}
\label{eq:51}
L_{1}=\sqrt{\gamma}\ \sigma_{y}.
\end{equation}
Using $\left\langle 0\right|L_{1}\left|0\right\rangle = \left\langle 1\right|L_{1}\left|1\right\rangle = 0$ and $\left\langle 0\right|L_{1}\left|1\right\rangle = -\left\langle 1\right|L_{1}\left|0\right\rangle = -i\sqrt{\gamma}$, we find $\dot{p}_{0}$ and $\dot{p}_{1}$ as 
\begin{equation}
\begin{aligned}
\dot{p}_{0} &= \gamma(p_{1} -p_{0})\\
\dot{p}_{1} &= \gamma(p_{0} - p_{1}).
\end{aligned}
\label{eq:52}
\end{equation}
This result looks like the bit flip channel given by Eq. (\ref{eq:46}) since the phase flip process does not affect the probability according to the result found in Eq. (\ref{eq:49}).  To verify Eq. (\ref{eq:9}), using the same observable $A=J_{z}$, we can write
\begin{equation}
\begin{aligned}
\left\langle A\right\rangle =Tr\left( \rho A\right) &= \frac{1}{2} \left( p_{0}-p_{1}\right) \\
\left\langle A^{2}\right\rangle  &= Tr\left( \rho A^{2}\right) =\frac{1}{4}\left( p_{0}+p_{1}\right) =\frac{1}{4} \\
\sigma_{A} &= \sqrt{p_{0}p_{1}},
\end{aligned}
\label{eq:52.1} 
\end{equation}
which can be used to find

\begin{equation}
\label{eq:52.2}
\dot{\sigma}_{A}=\frac{\gamma \left( p_{1}-p_{0}\right) ^{2}}{2\sqrt{p_{0}p_{1}}},
\end{equation}
and obtain
\begin{equation}
\label{eq:52.3}
2\sigma_{A}\dot{\sigma}_{A} = \gamma \left( p_{1}-p_{0}\right) ^{2}.
\end{equation}
Note that unlike \Eq{eq:40.2}, the term $2\,\sigma_A$ does not factor out in the final expression in
\Eq{eq:52.3}.
To find the first term on the RHS of Eq. (\ref{eq:9}), we can use $\Delta A=A-\left\langle A\right\rangle  I$ to write
\begin{equation}
\label{eq:52.4}
 \left\langle 0\right|(\Delta A)^{2}\left|0\right\rangle  = p^{2}_{1} \qquad  \left\langle 1\right|(\Delta A)^{2}\left|1 \right\rangle  = p^{2}_{0},
\end{equation}
which yields
\begin{equation}
\label{eq:52.5}
\sum_{j}\dot{p}_{j}\left\langle (\Delta A)^{2}\right\rangle_{j} = \dot{p}_{0} p^{2}_{1} + \dot{p}_{1}p^{2}_{0} = \gamma \left( p_{1}-p_{0}\right) ^{2}. 
\end{equation}
The results from Eqs. (\ref{eq:52.3}, \ref{eq:52.5}) indicate that
\begin{equation}
\label{eq:52.6}
2\sigma_{A}\dot{\sigma}_{A} = \sum_{j}\dot{p}_{j}\left\langle (\Delta A)^{2}\right\rangle_{j} = \gamma \left( p_{1} - p_{0}\right) ^{2} .
\end{equation}
The result found in Eq. (\ref{eq:52.6}) indicates that since $\gamma>0$ the variance in nondecreasing as long as $p_0\neq p_1$.  Unlike the result in amplitude damping example, where the variance could increase or decrease due to  population imbalance in the ground state or the excited state, bit flip and bit-phase flip can produce a nonnegative rate of change of the variance proportional to the imbalance squared except for the case where the populations are equal.

\subsection{Superradiance}

For our final example we would like to consider superradiance process and compute $\dot{p}_{0}$ and $\dot{p}_{1}$.  In the superradiance process $N$ identical systems interact collectively with the environment.  Since the systems couple to their common environment, the spontaneous emission due to each system is not independent any more.  Rather, they emit cooperatively, producing an intense burst of radiation with peak intensity as a function of $N^{2}$ instead of $N$.  The collective decay operator is given by \cite{BP}
\begin{equation}
\label{eq:53}
J_{-}= \sum_{i=1}^{N} \sigma^{(i)}_{-}
\end{equation}   
where $\sigma^{(i)}_{-}$ lowers the $i$th atom.  Then one can write
\begin{equation}
\label{eq:54}
\frac{d\rho}{dt}= - \frac{i}{\hbar} \left[ H,\rho\right]  + \gamma \left( 2 J_{-}\rho J_{+} - \left\lbrace J_{+}J_{-} , \rho\right\rbrace\right),  
\end{equation}
where $J_{+}=J^{\dagger}_{-}$.  Comparing Eq. (\ref{eq:54}) with the general time evolution of the density operator in an open quantum system given by
\begin{equation}
\label{eq:55}
\frac{d\rho}{dt}= - \frac{i}{\hbar} \left[ H,\rho\right]  + \sum_{k} \left( L_{k}\rho L^{\dagger}_{k} - \frac{1}{2}\left\lbrace L^{\dagger}_{k}L_{k} , \rho\right\rbrace\right)
\end{equation}
we can write the Lindblad operator as
\begin{equation}
\label{eq:56}
L=\sqrt{2 \gamma} \ J_{-},
\end{equation}
which can be used to find the corresponding Kraus operator as
\begin{equation}
\label{eq:57}
E_{1}=\sqrt{dt} \ L_{1} = \sqrt{2 \gamma dt} \ J_{-}.
\end{equation}
 We now consider a single two-level system where the ground system and the first excited state are given by $\left|0\right\rangle $ and $\left|1\right\rangle $, respectively.  Knowing that $\sigma_{-} \left|0\right\rangle =0$ and $\sigma_{-}\left|1\right\rangle =\left|0\right\rangle $ one can write $\sigma_{-}$ as
\begin{equation}
\label{eq:58}
\sigma_{-} = \begin{pmatrix}
0 & 1\\
0 & 0
\end{pmatrix} 
\end{equation}
and knowing that $\sigma_{+}=\sigma^{\dagger}_{-}$, we can write 
\begin{equation}
\label{eq:59}
\sigma_{+} = \begin{pmatrix}
0 & 0\\
1 & 0
\end{pmatrix}. 
\end{equation}
Considering $N=2$ we can find the collective decay operator as
\begin{equation}
\label{eq:60}
J_{-}=\sigma^{(1)}_{-} + \sigma^{(2)}_{-} = \sigma_{-} \otimes I + I \otimes \sigma_{-} = \begin{pmatrix}
0 & 1\\
0 & 0
\end{pmatrix} 
\otimes \begin{pmatrix}
1 & 0\\
0 & 1
\end{pmatrix}
+ \begin{pmatrix}
1 & 0\\
0 & 1
\end{pmatrix}
\otimes \begin{pmatrix}
0 & 1\\
0 & 0
\end{pmatrix}   
\end{equation}
where $\sigma^{(1)}_{-}=\sigma_{-} \otimes I$ gives $\sigma_{-}$ acting on atom 1, and $\sigma^{(2)}_{-}=I \otimes \sigma_{-}$ gives $\sigma_{-}$ acting on atom 2.  The result found in Eq. (\ref{eq:60}) gives the collective decay operator as
\begin{equation}
\label{eq:61}
J_{-}= \begin{pmatrix}
0 & 1 & 1 & 0\\
0 & 0 & 0 & 1\\
0 & 0 & 0 & 1\\
0 & 0 & 0 & 0
\end{pmatrix}.
\end{equation}
We can then compute the result of $\sigma_{-}$ on each state as
\begin{equation}
\begin{aligned}
J_{-}\left|00\right\rangle &= \left( \sigma^{(1)}_{-}+\sigma^{(2)}_{-}\right) \left|00\right\rangle =0\\
J_{-}\left|01\right\rangle &= \left|00\right\rangle \\
J_{-}\left|10\right\rangle &= \left|00\right\rangle \\
J_{-}\left|11\right\rangle &= \left|01\right\rangle  + \left|10\right\rangle .
\end{aligned}
\label{eq:62}
\end{equation}
Using Eq. (\ref{eq:56}) we find the nonzero elements as
\begin{equation}
\label{eq:63}
\left\langle 00\right|L\left|01\right\rangle = \left\langle 00\right|L\left|10\right\rangle = \left\langle 01\right|L\left|11\right\rangle = \left\langle 10\right|L\left|11\right\rangle = \sqrt{2 \gamma},
\end{equation}
which can be used to find $\dot{p}_{00}$ and $\dot{p}_{11}$ as
\begin{equation}
\begin{aligned}
\dot{p}_{00} &= p_{01} \left|\left\langle 00\right|L\left|01\right\rangle \right|^{2} + p_{10}\left|\left\langle 00\right|L\left|10\right\rangle \right|^{2} = 2\gamma (p_{01}+p_{10})\\
\dot{p}_{11} &= - p_{11} \left|\left\langle 11\right|L\left|10\right\rangle \right|^{2} - p_{11}\left|\left\langle 11\right|L\left|01\right\rangle \right|^{2} = -4\gamma p_{11}.
\end{aligned}
\label{eq:64}
\end{equation}
The result found in Eq. (\ref{eq:64}) indicate that the state in which both atoms are in the excited state loses probability as photon is being emitted at the rate $4\gamma$.  Furthermore, probability flows from the state $\left| 11\right\rangle $ to $\left|01\right\rangle $ and $\left|10\right\rangle $ and eventually to the ground state $\left|00\right\rangle $.  For completeness, we can find $\dot{p}_{10}$ and $\dot{p}_{01}$ and see that the time derivative of the sum of all probabilities is zero.
To find $\dot{p}_{10}$ and $\dot{p}_{01}$ we have
\begin{equation}
\begin{aligned}
\dot{p}_{10} &= p_{11}\left|\left\langle 10\right|L\left|11\right\rangle \right|^{2} - p_{10} \left|\left\langle 00\right|L\left|10\right\rangle \right|^{2}=2\gamma (p_{11}-p_{10}) \\
\dot{p}_{01} &= p_{11}\left|\left\langle 01\right|L\left|11\right\rangle \right|^{2} - p_{01}\left|\left\langle 00\right|L\left|01\right\rangle \right|^{2} = 2\gamma (p_{11}-p_{01})
\end{aligned}.
\label{eq:65}
\end{equation}
Using Eqs. (\ref{eq:64}, \ref{eq:65}), it is easy to see that
\begin{equation}
\label{eq:66}
\dot{p}_{00}+\dot{p}_{10}+\dot{p}_{01}+\dot{p}_{11}=0,
\end{equation}
as expected.  To verify Eq. (\ref{eq:9}), use observable $A=J_{z}$ that is given by  
\begin{equation}
\label{eq:66.1}
J_{z} = \frac{1}{2}
\begin{pmatrix}
1 & 0 & 0 & 0 \\
0 & 0 & 0 & 0 \\
0 & 0 & 0 & 0 \\
0 & 0 & 0 & -1
\end{pmatrix} .
\end{equation}
We can then find 
\begin{equation}
\begin{aligned}
\left\langle A\right\rangle &= Tr\left( \rho A\right) = p_{00}-p_{11}\\
\left\langle A^{2}\right\rangle  &= Tr\left( \rho A^{2}\right) =p_{00}+p_{11} \\
\sigma^{2}_{A} &= p_{00}+p_{11}-\left( p_{00}-p_{11}\right) ^{2},
\end{aligned}
\label{eq:66.2}
\end{equation}
which can be used to write
\begin{equation}
\label{eq:66.3}
\dot{\sigma}_{A}=\frac{\dot{p}_{00}+\dot{p}_{11}-2\left( \dot{p}_{00}-\dot{p}_{11}\right) \left( p_{00}-p_{11}\right) }{2\sqrt{p_{00}+p_{11}-\left( p_{00}-p_{11}\right) ^{2}}}.
\end{equation}
For the first term on the RHS of Eq. (\ref{eq:9}) we have
\begin{equation}
\begin{aligned}
\left\langle 00\right|(\Delta A)^{2}\left|00\right\rangle &= \left( 1+p_{11}-p_{00}\right) ^{2}\\
\left\langle 01\right|(\Delta A)^{2}\left|01\right\rangle &= \left( p_{11}-p_{00}\right) ^{2}\\
\left\langle 10\right|(\Delta A)^{2}\left|10\right\rangle &= \left( p_{11}-p_{00}\right) ^{2}\\
\left\langle 11\right|(\Delta A)^{2}\left|11\right\rangle &= \left( p_{11}-p_{00}-1 \right) ^{2}\\
\end{aligned},
\label{eq:66.4}
\end{equation}
which can be used to write
\begin{equation}
\begin{aligned}
\sum_{j}\dot{p}_{j}\left\langle (\Delta A)^{2}\right\rangle _{j} &= \dot{p}_{00}\left( 1+p_{11}-p_{00}\right) ^{2} + \dot{p}_{01}\left( p_{11}-p_{00}\right) ^{2} + \dot{p}_{10}\left( p_{11}-p_{00}\right) ^{2} +\dot{p}_{11} \left( p_{11}-p_{00}-1 \right) ^{2}\\
 &= \dot{p}_{00}+\dot{p}_{11} -2\left( p_{00}-p_{11}\right) \left( \dot{p}_{00}-\dot{p}_{11}\right) ,
\end{aligned}
\label{eq:66.5} 
\end{equation}
where we have used $\dot{p}_{00}+\dot{p}_{01}+\dot{p}_{10}+\dot{p}_{11}=0$ to get the last line of Eq. (\ref{eq:66.5}).  The results from Eqs. (\ref{eq:66.3}, \ref{eq:66.5}) yield
\begin{equation}
\label{eq:66.6}
2 \sigma_{A}\dot{\sigma}_{A}=\sum_{j}\dot{p}_{j}\left\langle (\Delta A)^{2}\right\rangle _{j} = \dot{p}_{00}+\dot{p}_{11} -2\left( p_{00}-p_{11}\right) \left( \dot{p}_{00}-\dot{p}_{11}\right).
\end{equation}

\begin{figure}[h]
\begin{center}
\begin{tabular}{ccc}
\includegraphics[width=3.0in,height=1.75in]{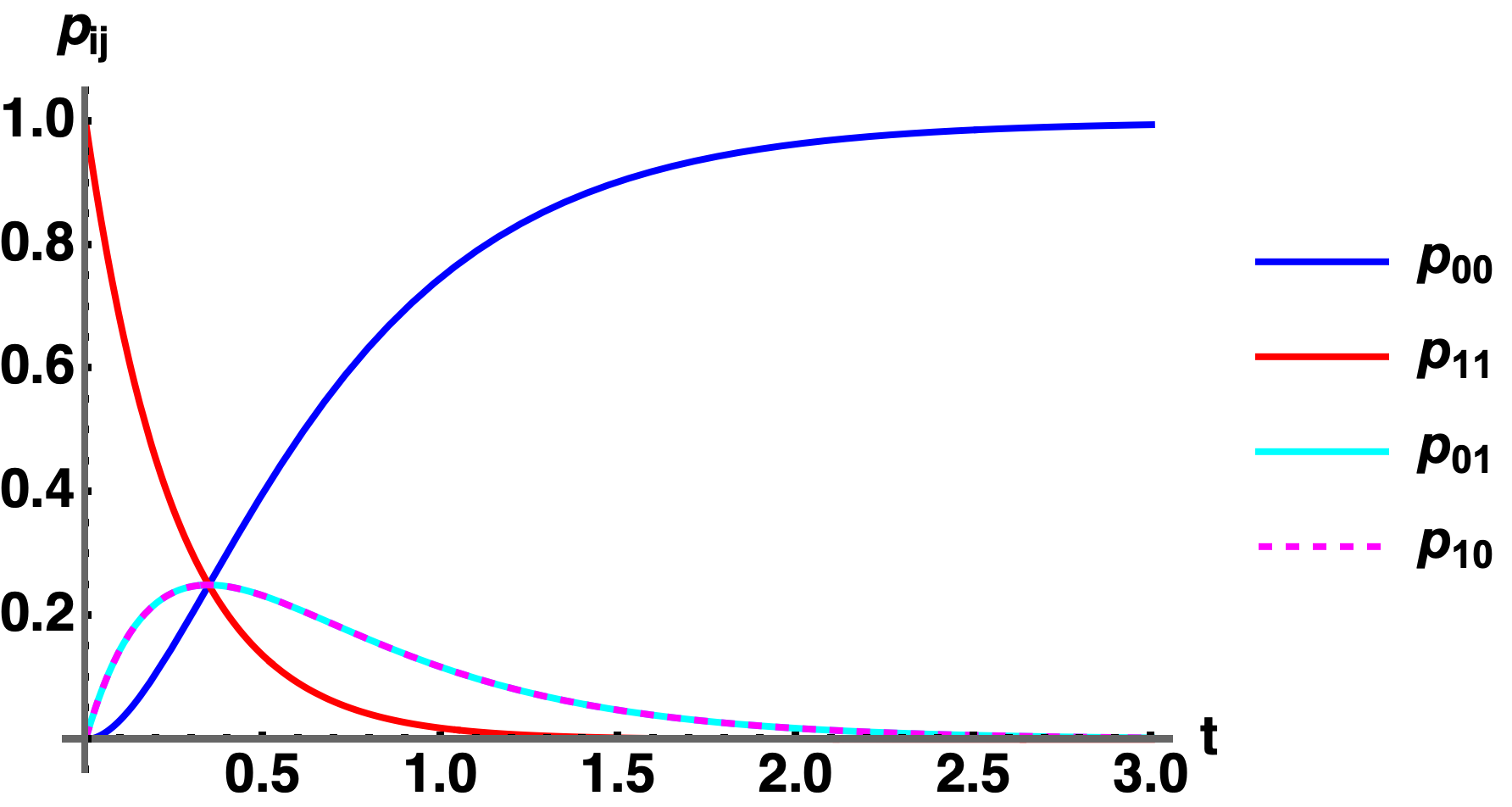} \hspace{0.25in} &
{}&
\includegraphics[width=3.0in,height=1.75in]{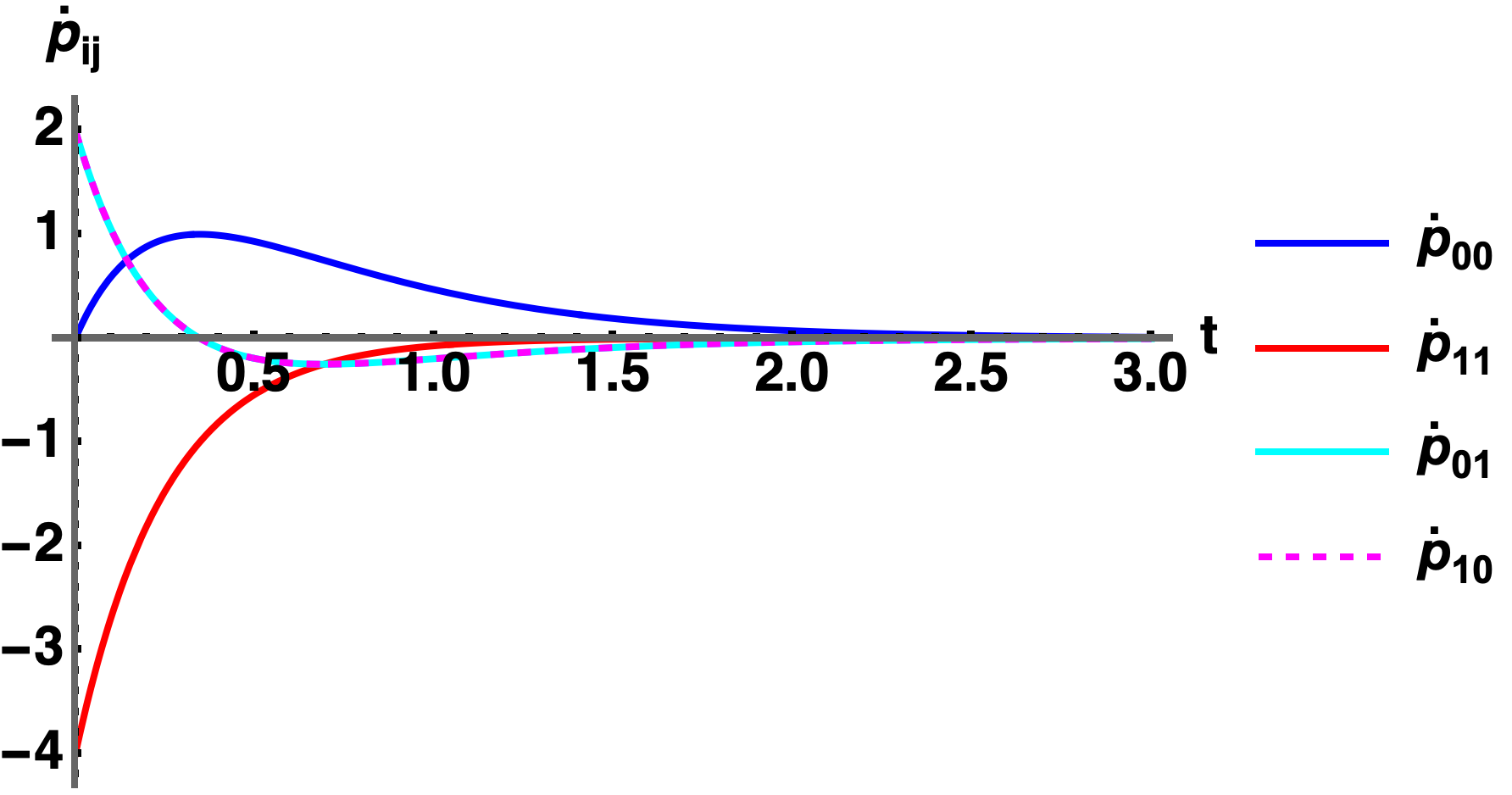}
\end{tabular}
\end{center}
\caption{
(left)  $p_{ij}(\tau)$ and 
(right) $\dot{p}_{ij}(\tau)$
vs $\tau = \gamma\,t$.
}\label{fig:p:pdot:solns}
\end{figure}
\begin{figure}[h]
\begin{center}
\includegraphics[width=3.25in,height=1.75in]{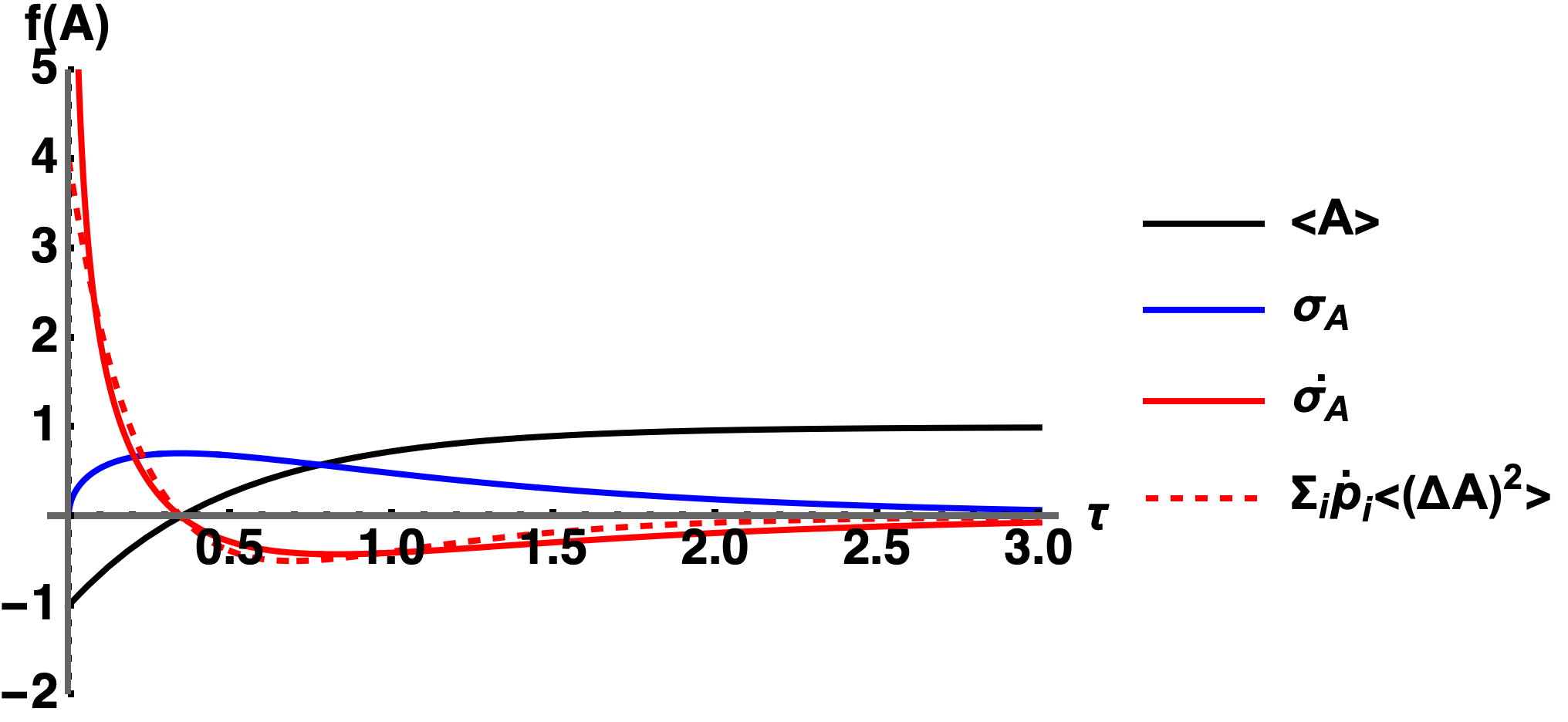} 
\end{center}
\caption{ $\avg{A},\, \sigma_A,\, \dot{\sigma}_A,\; \trm{and}\; \sum_i \dot{p}_i \avg{(\DA)^2}$
vs $\tau = \gamma\,t$, for $A = J_z$.
}\label{fig:f:A}
\end{figure}
We can easily solve the above equations for the above probabilities $p_{ij}(\tau)$ and their derivatives  $\dot{p}_{ij}(\tau)$,  where $\tau = \gamma t$:
\bea{p:pdot:solns}
\big[ p_{00}(\tau),\, p_{11}(\tau),\, p_{01}(\tau),\, p_{10}(\tau) \big] 
&=& 
\left[e^{-4 \tau } \left(e^{2 \tau
   }-1\right)^2,e^{-4 \tau },e^{-4
   \tau } \left(e^{2 \tau
   }-1\right),e^{-4 \tau } \left(e^{2
   \tau }-1\right)\right], \\
\big[ \dot{p}_{00}(\tau),\, \dot{p}_{11}(\tau),\, \dot{p}_{01}(\tau),\, \dot{p}_{10}(\tau) \big]
&=&   
\left[4 e^{-4 \tau } \left(e^{2 \tau
   }-1\right),-4 e^{-4 \tau },-2 e^{-4
   \tau } \left(e^{2 \tau
   }-2\right),-2 e^{-4 \tau }
   \left(e^{2 \tau }-2\right)\right], 
\eea
from which we can form the quantities for $A=J_z$:
\be{f:A}
\Big[
\avg{A},\, \sigma_A,\, \dot{\sigma}_A,\, \sum_i \dot{p}_i \avg{(\DA)^2}
\Big] 
= 
\Big[1-2 e^{-2 \tau },\sqrt{2} e^{-2
   \tau } \sqrt{e^{2 \tau
   }-1},-\frac{\sqrt{2} e^{-2 \tau }
   \left(e^{2 \tau
   }-2\right)}{\sqrt{e^{2 \tau
   }-1}},-4 e^{-4 \tau } \left(e^{2
   \tau }-2\right)\Big],
\ee
which we plot in \Fig{fig:p:pdot:solns} and \Fig{fig:f:A} above.
As in the case of \Eq{eq:52.3}, the term $2\,\sigma_A$ does not factor out of the final expression in \Eq{eq:66.3}.
In fact, since $\sigma_A(0)=0$, as seen from \Eq{f:A}, we have that 
$\dot{\sigma}_A~=~\sum_i \dot{p}_i \avg{(\DA)^2}/(2\, \sigma_A)\overset{\tau\to 0}{\to}~\infty$,
although the non-unitary contribution $\sum_i \dot{p}_i \avg{(\DA)^2}$ remains finite at $\tau=0$.
This is a generic feature of $\dot{\sigma}_{A}$, if one begins in an initial state of zero fluctuations
$\sigma_A(0)=0$.
\bk

According to Eq. (\ref{eq:66.6}) the change in the variance depends on the effects due to the interaction of the two two-level systems with a radiation field. The first two terms describe the change in the combined population of the states $p_{00}$ and $p_{11}$ , while the last term accounts for the contribution of the changing population imbalance.  Consequently, the variance dynamics depend not only on the populations themselves but also on how they  change with time.

While the examples discussed do not indicate any 
unexpected results, exploring  them is worthwhile as they provide a clear demonstration of the results found for the time evolution of the eigenvalue of the density operator in terms of the Kraus and the Lindblad operators.  Furthermore, these examples can show how the response of the uncertainty of the observable can be used to distinguish different noise mechanisms in a two level quantum system, for the bit-phase flip and the amplitude damping examples.  While the superradiance example provides a way to investigate how collective emission redistributes populations and modifies fluctuations.   

\section{Conclusion}

In this work, we investigated the time evolution of the uncertainty of a quantum observable and established connections between fluctuation dynamics, Fisher information, and the quantum operations governing open-system evolution. Starting from Eq. (\ref{eq:9}), $
2\sigma_A\dot{\sigma}_A
=
\sum_j\dot{p}_j\langle(\Delta A)^2\rangle_j
+
\left\langle\{\Delta A,\dot A\}\right\rangle$ ,
we derived upper bounds on the rate of fluctuation growth using the Cauchy–Schwarz inequality.  For closed systems, the resulting bound reduces to the known relation involving the variance of $\dot{A}$.

We next examined Eq. (\ref{eq:9}) from an information-theoretic perspective.  By introducing the symmetric logarithmic derivative (SLD) and separating the incoherent and coherent contributions to the density operator, we related the evolution of fluctuations of observable $A$ to the corresponding incoherent and coherent components of the quantum Fisher information (QFI). The resulting bound shows that the growth of fluctuations is constrained by the information content of the evolving quantum state, as well as the explicit time dependence of the observable.  Thus separating the contributions from the environment, the Hamiltonian, and the observable itself to changes in the dynamics of fluctuations of $A$.

We then investigated the relation between Eq. (\ref{eq:9}) and quantum operator.  By expressing the density operator in its eigenbasis, we derived a general expression for the time derivative of its eigenvalues in terms of Kraus operators.  This result provides a direct connection between the change in the statistical weights of the quantum state and the quantum channel responsible for its evolution.  Using the infinitesimal relation between Kraus and Lindblad operators, we further obtained a representation of $\dot{p}_{i}$ in terms of Lindblad operators. Consequently, the contribution involving $\dot{p}_{i}$ in Eq. (\ref{eq:9}) which expresses how the eigenvalues (populations) of the density matrix change in time, can be related directly to the physical processes causing that change.

The examples considered demonstrate the utility of this formulation. Amplitude damping and bit-flip or bit-phase-flip processes produce population transfer between states, whereas the phase-flip channel leaves the populations unchanged while modifying phase information.  The collective superradiance example further illustrates how the structure of a Lindblad operator can produce enhanced transition rates and correlated population dynamics.  In each case, the derived expressions reproduce the expected probability flow and preserve the normalization of the density operator.  

Overall, our results provide a unified framework for understanding fluctuation dynamics in open quantum systems. By linking the rate of change of an observable’s uncertainty to quantum Fisher information, state populations, and Kraus–Lindblad descriptions of quantum channels, we clarify both the informational and physical origins of fluctuation growth. This approach may be useful for characterizing quantum noise, identifying dynamical resources in nonequilibrium processes, and designing control strategies for quantum sensing, communication, and information-processing platforms.

\bigskip


\appendix
\section{}\label{app:A}

In this Appendix we offer a derivation of Eq. (\ref{eq:b1}).   
We start by rewriting Eq. (\ref{eq:9}) as
\begin{equation}
\begin{aligned}
2\sigma_{A}\dot{\sigma}_{A} &= Tr\left( \dot{\rho} (\Delta A)^{2}\right) - \sum_{j} p_{j} \left( \left\langle \dot{\psi}_{j}\right| (\Delta A)^{2}\left|\psi_{j}\right\rangle +\left\langle \psi_{j}\right|(\Delta A)^{2}\left|\dot{\psi}_{j}\right\rangle \right) +\left\langle \left\lbrace \Delta A,\dot{A}\right\rbrace \right\rangle \\
&= Tr\left( \dot{\rho} (\Delta A)^{2}\right) - i \sum_{j}p_{j} \left( \left\langle \psi_{j}\right| \left( H(\Delta A)^{2} - (\Delta A)^{2}H\right) \left|\psi_{j}\right\rangle \right) +\left\langle \left\lbrace \Delta A,\dot{A}\right\rbrace \right\rangle\\
&= Tr\left( \dot{\rho} (\Delta A)^{2}\right) -i \left\langle \left[ H,(\Delta A)^{2}\right] \right\rangle +\left\langle \left\lbrace \Delta A,\dot{A}\right\rbrace \right\rangle .
\end{aligned}
\label{eq:A1}
\end{equation}
Moreover, we know that
\begin{equation}
\label{eq:A2}
\left[ H,(\Delta A)^{2}\right]= \left\lbrace \Delta A, [H,\Delta A]\right\rbrace = \left\lbrace \Delta A, [H,A]\right\rbrace ,  
\end{equation}
where we have used $[H,\Delta A]=[H,A]-[H,I]\left\langle A\right\rangle = [H,A]$, which allows us to rewrite Eq. (\ref{eq:A1}) as
\begin{equation}
\label{eq:A3}
2\sigma_{A}\dot{\sigma}_{A}=Tr\left( \dot{\rho} (\Delta A)^{2}\right)- i\left\langle \left\lbrace \Delta A,[H,A]\right\rbrace \right\rangle +\left\langle \left\lbrace \Delta A, \dot{A}\right\rbrace \right\rangle = Tr\left( \dot{\rho} (\Delta A)^{2}\right)+\left\langle \left\lbrace \Delta A, \dot{A}-i[H,A]\right\rbrace \right\rangle  .
\end{equation}
Using $\dot{A}=i[H,A]+\partial_{t}A$, we can rewrite Eq. (\ref{eq:A3}) as
\begin{equation}
\label{eq:A3.1}
2\sigma_{A}\dot{\sigma}_{A}=Tr\left( \dot{\rho} (\Delta A)^{2}\right)+ \left\langle \left\lbrace \Delta A,\partial_{t}A\right\rbrace \right\rangle .
\end{equation}
This concludes our derivation of Eq. (\ref{eq:b1}).

\section{}\label{app:B}
In this Appendix we include a derivation of Eq. (\ref{eq:37}).
To achieve this, use Eq. (\ref{eq:28}) to write $E_{0}\rho E^{\dagger}_{0}$ as 
\begin{equation}
\begin{aligned}
E_{0}\rho E^{\dagger}_{0} &= \left(I-i\Omega \mathrm{d}t-\frac{1}{2}\sum_{k}L^{\dagger}_{k}L_{k} \mathrm{d}t\right) \rho \left( I+i\Omega \mathrm{d}t-\frac{1}{2}\sum_{k}L^{\dagger}_{k}L_{k} dt\right) \\ 
 & \simeq \rho -i\left[ \Omega , \rho\right] \mathrm{d}t - \frac{1}{2} \sum_{k} \left\lbrace L^{\dagger}_{k}L_{k} , \rho\right\rbrace  \mathrm{d}t,
\label{eq:B31}
\end{aligned}
\end{equation}
where we have ignored higher order terms for an infinitesimal time step.
Projecting the result found in Eq. (\ref{eq:B31}) onto $\left|\psi_{i}\right\rangle $  yields
\begin{equation}
\label{eq:B32}
\left\langle \psi_{i}\right|E_{0}\rho E^{\dagger}_{0} \left|\psi_{i}\right\rangle = \left\langle \psi_{i}\right|\rho \left|\psi_{i}\right\rangle -i \mathrm{d}t \left\langle \psi_{i}\right|\left[ \Omega,\rho\right]  \left|\psi_{i}\right\rangle-\frac{dt}{2} \sum_{k}\left\langle \psi_{i}\right|\left\lbrace L^{\dagger}_{k}L_{k},\rho\right\rbrace \left|\psi_{i}\right\rangle , 
\end{equation}
which using
\begin{equation}
\label{eq:B33}
\left\langle \psi_{i}\right|\Omega \rho \left|\psi_{i}\right\rangle -\left\langle \psi_{i}\right|\rho \Omega \left|\psi_{i}\right\rangle =  p_{i}\left\langle \psi_{i}\right| \Omega \left|\psi_{i}\right\rangle - p_{i}\left\langle \psi_{i}\right| \Omega \left|\psi_{i}\right\rangle = 0 ,
\end{equation}
becomes
\begin{equation}
\label{eq:B34}
\left\langle \psi_{i}\right|E_{0}\rho E^{\dagger}_{0} \left|\psi_{i}\right\rangle = p_{i} - \frac{\mathrm{d}t}{2} \sum_{k}\left\langle \psi_{i}\right|\left\lbrace L^{\dagger}_{k}L_{k},\rho\right\rbrace \left|\psi_{i}\right\rangle .
\end{equation}
The second term on the RHS of Eq. (\ref{eq:B34}) can be written as
\begin{equation}
\begin{aligned}
-\frac{\mathrm{d}t}{2}\left\langle \psi_{i} \right|\left\lbrace L^{\dagger}_{k}L_{k},\rho\right\rbrace \left|\psi_{i}\right\rangle  &=
-\frac{\mathrm{d}t}{2}\left( \left\langle \psi_{i}\right| L^{\dagger}_{k}L_{k}\rho \left|\psi_{i}\right\rangle +\left\langle \psi_{i}\right| \rho L^{\dagger}_{k}L_{k} \left|\psi_{i}\right\rangle\right)  \\
&= -\frac{dt}{2} \left( p_{i}\left\langle \psi_{i}\right|L^{\dagger}_{k}L_{k}\left|\psi_{i}\right\rangle + p_{i}\left\langle \psi_{i}\right| L^{\dagger}_{k}L_{k}\left|\psi_{i}\right\rangle \right)\\
&= -p_{i} dt \left\langle \psi_{i}\right| L^{\dagger}_{k}L_{k}\left| \psi_{i}\right\rangle \\
&= -p_{i} dt\left\langle \psi_{i}\right| L^{\dagger}_{k}\left( \sum_{j}\left|\psi_{j}\right\rangle \left\langle \psi_{j}\right|\right) L_{k}\left|\psi_{i}\right\rangle \\
&= - p_{i} \mathrm{d}t \sum_{j}\left|\left\langle \psi_{j}\right| L_{k}\left|\psi_{i}\right\rangle \right|^{2} ,
\label{eq:B35}
\end{aligned}
\end{equation}
where in line 4 we have used $I=\sum_{j}\left|\psi_{j}\right\rangle \left\langle \psi_{j}\right|$, and in the last line we have used the fact that $\left\langle \psi_{i}\right|L^{\dagger}_{k}\left|\psi_{j}\right\rangle =\left\langle \psi_{j}\right|L_{k}\left|\psi_{i}\right\rangle ^{*}$.

Next, considering the dissipative terms given by $E_{k}$ in Eq. (\ref{eq:28}), we have
\begin{equation}
\begin{aligned}
\left\langle \psi_{i}\right|E_{k}\rho E^{\dagger}_{k}\left|\psi_{i}\right\rangle &= \mathrm{d}t\left\langle \psi_{i}\right| L_{k}\rho L^{\dagger}_{k}\left|\psi_{i}\right\rangle \\
&= \mathrm{d}t\sum_{j}p_{j}\left\langle \psi_{i}\right|L_{k}\left|\psi_{j}\right\rangle \left\langle \psi_{j}\right|L^{\dagger}_{k}\left|\psi_{i}\right\rangle \\
&= \mathrm{d}t\sum_{j}p_{j}\left|\left\langle \psi_{i}\right|L_{k}\left|\psi_{j}\right\rangle \right|^{2},
\label{eq:B36}
\end{aligned}
\end{equation}
where in the second line we have used $\rho=\sum_{j}p_{j}\left|\psi_{j}\right\rangle \left\langle \psi_{j}\right|$.  Upon substituting the results found in Eqs.(\ref{eq:B34}-\ref{eq:B36}) into Eq. (\ref{eq:30}), we can now obtain $\dot{p}_{i}$ from Eq. (\ref{eq:19}) as 
\begin{equation}
\begin{aligned}
\dot{p}_{i} &= \lim_{\mathrm{d}t \to 0}\frac{1}{\mathrm{d}t}\left( p_{i}(t+\mathrm{d}t)-p_{i}(t)\right) =\lim_{\mathrm{d}t \to 0}\frac{1}{dt} \left(\sum_{k}\left\langle \psi_{i}\right|E^{\dagger}_{k}\rho E_{k} \left|\psi_{i}\right\rangle +\left\langle \psi_{i}\right|E_{0}\rho E^{\dagger}_{0}\left|\psi_{i}\right\rangle -p_{i}\right)\\ 
&= \lim_{\mathrm{d}t \to 0}\frac{1}{\mathrm{d}t} \left( dt \sum_{kj}p_{j}\left|\left\langle \psi_{i}\right|L_{k}\left|\psi_{j}\right\rangle \right|^{2}+p_{i} - p_{i} \mathrm{d}t \sum_{kj}\left|\left\langle \psi_{j}\right|L_{k}\left|\psi_{i}\right\rangle\right|^{2}-p_{i}\right) \\ 
&=\sum_{kj} \left( p_{j}\left|\left\langle \psi_{i}\right|L_{k}\left|\psi_{j}\right\rangle \right|^{2}-p_{i}\left|\left\langle \psi_{j}\right|L_{k}\left|\psi_{i}\right\rangle \right|^{2}\right) .
\label{eq:B37}
\end{aligned}
\end{equation}
With this result we conclude our derivation of Eq. (\ref{eq:37}).

\section{}\label{app:C}
In this Appendix, we wish to show directly that 
$\cov(L_I A_I ) = \Tr[(\dot{\rho})_I A_I]$, and
$\cov(L_C A_C ) = \Tr[(\dot{\rho})_C A_C]$.

Using the definitions
\be{}
L_I \equiv \sum_j \frac{d\ln p_j}{dt} |j\rangle\langle j|, \quad\trm{and}\quad
L_C \equiv (2i) \sum_{j \neq k} \frac{\langle j | [\rho, H] | k \rangle}{(p_j + p_k)} |j\rangle\langle k| 
\ee
we can compute the following.

\subsection*{The Incoherent part}
Now, $\Tr[(\dot{\rho})_I A_I]$ is simple to show since all operators are diagonal.
\begin{align}
\avg{L_I A_I} = \avg{A_I L_I }  = \Tr[\rho\, L_I A_I] & = 
\Tr[\sum_i p_i \ket{i}\bra{i} \, \sum_j \frac{\dot{p}_j}{p_j} \ket{j}\bra{j} \, \sum_k A_k \ket{k}\bra{k}] 
= \sum_i \dot{p}_i A_i \equiv\Tr[(\dot{\rho})_I A_I]\\
\implies \trm{\cov}(L_I A_I) &= \half \avg{L_I A_I + A_I L_I}-\avg{L_I}\ev{A_I}\equiv \avg{L_I A_I} = Tr[(\dot{\rho})_I A_I].
\end{align}
since $\avg{L_I} = \Tr[\rho \sum_i \frac{\dot{p}_i}{p_i} \ket{i}\bra{i}] 
= \sum_i \dot{p}_i = \frac{d}{dt}\big(\sum_i p_i=1\big)=0$.

\subsection*{The Coherent part}
The coherent part takes a little bit more work, as shown in \cite{Gor} (Appendix C).
We want to show that\newline $\trm{cov}(L_C A_C) = Tr[(\dot{\rho})_C A_C]$.

Consider first 
\begin{align}
\ev{A_C L_C}=\Tr[\rho A_C L_C] &= \operatorname{Tr}\left[ \rho \sum_{j \neq k} A_{jk} |j\rangle\langle k| \sum_{j' \neq k'} (2i) \frac{\langle j' | [\rho, H] | k' \rangle}{p_{j'} + p_{k'}} |j'\rangle\langle k'| \right] \\
&= (2i)\,\sum_{j \neq k, j' \neq k'}  
\underbrace{\langle k' | \rho | j \rangle}_{p_j \delta_{k' j}}  \delta_{k j'}
A_{jk}  \frac{\langle j' | [\rho, H] | k' \rangle}{p_{j'} + p_{k'}} \\
& = (2i) \sum_{j \neq k} p_j A_{jk} \frac{\langle k | [\rho, H] | j \rangle}{p_j + p_k} 
\end{align}

Similarly
\begin{align}
\ev{L_C A_C} = 
\Tr[\rho L_C A_C] &= \operatorname{Tr}\left[ \rho \sum_{j \neq k} (2i) \frac{\langle j | [\rho, H] | k \rangle}{p_j + p_k} |j\rangle\langle k| \sum_{j' \neq k'} A_{j'k'} |j'\rangle\langle k'| \right] \\
&= (2i) \sum_{j \neq k, j' \neq k'} 
\underbrace{\langle k' | \rho | j \rangle}_{p_j \delta_{k'j}} 
\frac{\langle j | [\rho, H] | k \rangle}{p_j + p_k} A_{j'k'} \delta_{k j'} \\
&= (2i) \sum_{j \neq k} p_j \frac{\langle j | [\rho, H] | k \rangle}{p_j + p_k} A_{kj} 
\overset{j\leftrightarrow k}{=} 
(2i) \sum_{j \neq k} p_k A_{jk} \frac{\langle k | [\rho, H] | j \rangle}{p_j + p_k} 
\end{align}

Adding $\half [(C6) + (C9)]$ we have:
\be{}
\hspace{-0.5in}
\frac{1}{2}\ev{ \{L_C, A_C\}}= \frac{(2i)}{2} \sum_{j \neq k} A_{jk} 
\frac{(p_j + p_k)}{(p_j + p_k)} 
\langle k | [\rho, H] | j \rangle = i \sum_{j \neq k} A_{jk} \langle k | [\rho, H] | j \rangle =
i\,\Tr\big[A_C [\rho,H]\, \big] \equiv  \operatorname{Tr}[(\dot{\rho})_C A_C] 
\ee

Finally, we can also add $\ev{L_C}\,\ev{A_C}$ since $\ev{L_C}=0$, as we show next:
\begin{align}
\langle L_C \rangle = \operatorname{Tr}[\rho L_C] &= (2i) \operatorname{Tr}\left[ \rho \sum_{j \neq k} \frac{\langle j | [\rho, H] | k \rangle}{p_j + p_k} |j\rangle\langle k| \right] \\
&= (2i) \sum_{j \neq k} \langle k | \rho | j \rangle \frac{\langle j | [\rho, H] | k \rangle}{p_j + p_k} \\
&= (2i) \sum_{j \neq k} p_j 
\underbrace{\delta_{kj}}_{=0} 
\frac{\langle j | [\rho, H] | k \rangle}{p_j + p_k} = 0 \quad (\text{since } j \neq k) 
\end{align}
Thus we can subtract $\langle L_C \rangle \langle A_C \rangle = 0$ to (C10)
to finally obtain:
\begin{align}
\boxed{\operatorname{cov}(L_C, A_C) = \frac{1}{2} \langle L_C A_C + A_C L_C \rangle - \langle L_C \rangle \langle A_C \rangle = \operatorname{Tr}[L\rho A_C] \equiv \Tr[(\dot{\rho})_C A_C]} 
\end{align}
Recall that in terms of the $\{p_i\}$ we had
\be{}
 i \sum_{j \neq k} (p_j - p_k) H_{jk} A_{kj} = i \sum_{j \neq k} A_{jk} \langle k | [\rho, H] | j \rangle = i \Tr\big[A_C \,[\rho, H]\big] \equiv \Tr[(\dot{\rho})_C A_C]
\ee

\end{document}